\documentclass[aps,ams,12pt,onecolumn,preprint,nofootinbib,showpacs,superscriptaddress]{revtex4}

\usepackage{graphicx}\usepackage{epsfig}

\usepackage[psamsfonts]{amssymb}
\usepackage{amsfonts}\usepackage{amscd}
\usepackage{amsmath,amssymb}
\usepackage{cases,bold_greek}
\everymath{\displaystyle}

\newcommand{\ba}{\begin{eqnarray}}
\newcommand{\ea}{\end{eqnarray}}
 \def  \Li3 { {\rm {Li}_3}  }
 \def  \li2  { { \rm {Li}_2 } }
\def  \poly4  { { \rm {Li}_4 } }
\def \Imm { \mbox{\rm Im} }
\def \Ree { \mbox{\rm Re} }

\graphicspath{{./Figures-2022-3_loop/}}

\usepackage{graphicx}\usepackage{epsfig}
\usepackage{amsfonts}\usepackage{amscd}
\usepackage{amsmath,amssymb}
\usepackage{cases,bold_greek}
\everymath{\displaystyle}
\begin{document}
\title{
 Lepton anomaly from QED diagrams  with vacuum  polarization \\ insertions
 within the Mellin-Barnes representation}
%Analytical calculations of the  radiative corrections to the
%\\ \noindent   lepton anomaly   up to the eighth  order
%from diagrams with \\ vacuum polarization insertions:
%dispersion relations  and \\ Mellin-Barnes techniques}
\author{O.P. Solovtsova}
\email{olsol@theor.jinr.ru ;solovtsova@gstu.gomel.by}
\affiliation{Bogoliubov Lab.
Theor. Phys., JINR, Dubna,
141980, Russia}
 \affiliation{Gomel State Technical
University, Gomel, 246746, Belarus}
\author{V.I. Lashkevich}
\affiliation{Gomel State
Technical University, Gomel, 246746, Belarus}
\author{L.P. Kaptari}
\email{kaptari@theor.jinr.ru} \affiliation{Bogoliubov Lab.
Theor. Phys., JINR, Dubna,
141980, Russia}

\begin{abstract}
The contributions to the anomalous magnetic moment of the lepton $L$
($L=e\ , \mu $ or  $\tau$) generated by a specific class of QED
diagrams are evaluated analytically up to the eighth order of the
electromagnetic coupling constant. The considered class of the Feynman
diagrams involves the vacuum polarization insertions into the
electromagnetic vertex of the lepton $L$ up to three closed lepton
loops. The corresponding analytical expressions are obtained as
functions of the mass ratios $r=m_l/m_L$ in the whole region $0 < r <
\infty$. Our consideration is based on a combined use of the
dispersion relations for the polarization operators and the
Mellin-Barnes integral transform for the Feynman parametric integrals.
This method is widely used in the literature in multi-loop  calculations
in relativistic quantum field theories. For each order of the
radiative correction, we derive analytical expressions as functions
of $r$, separately   at $r<1$ and $r>1$. We argue that  in spite of
the obtained explicit  expressions in these  intervals which are quite different, at first
glance, they represent  two branches of the same
analytical function. Consequently, for each order of  corrections
there is a unique analytical function defined in  the whole range of~$r\in (0,\infty)$.

The results of numerical calculations of the $4th$, $6th$ and $8th$ order corrections to the
anomalous magnetic moments of leptons  ($L=e ,\mu ,  \tau$) with all
possible vacuum polarization insertions   are represented as functions
of the ratio $r=m_l/m_L$. Whenever pertinent, we compare our
analytical expressions and  the corresponding asymptotical  expansions
with the known results available in the literature.
\end{abstract}


\maketitle

\thispagestyle{empty}
\date{\today}
%%%%%%%%%%%%%%%%%%%%%%%%%% Title %%%%%%%%%%%%%%%%%%%%%%%%%%%%%%%%%%%%%%
%\input bold_greek
\title{Some preliminary notes on the theory of the anomalous magnetic moment of a lepton: the electro-magnetic
contribution.}
\section{Introduction}

The giromagnetic factor $g$ is an important physical quantity relating
the magnetic-dipole moment of a particle to its spin. According to
Dirac's theory~\cite{dirac}, the   electron has $g=2$. The interaction
with photons shifts $g$, resulting in the famous notion of the  electron
anomaly, $a_e=(g_e-2)/2\neq 0$, which can be considered as a measure of
the magnetic field surrounding the electron. The effect  of shifts in
$g$ is also inherent in muons and tau leptons.  Albeit this anomaly is
rather small, its study  is of great importance since the systematic
discrepancy between the model-based theoretical calculations and the
experimentally measured giromagnetic factor   may indicate possible
limitations of the Standard Model (SM) or   possible existence of as yet
undiscovered particles  belonging to  SM,   however
  giving rise to some ``new physics''. Besides interactions with
virtual photons, the lepton giromagnetic factor  can acquire
corrections from hadronic vacuum polarization (HVP) and light by light
and electroweak scattering~\cite{review-2021,Jegerlehner:2017gek}.
Nowadays,   experimental  measurements of electron and muon
anomalies~\cite{Parker:2018vye,E989,E821} are performed with an
impressive precision, which allows for   meticulous comparison of data
with theoretical predictions, cf.
Refs.~\cite{Davoudiasl:2018fbb,Nomura,Keshavarzi:2020bfy,Malaescu}.

The pure QED contribution is by far the largest and can be calculated
numerically with an accuracy compatible with  the errors of
experimental measurements.  The  HVP  mechanism, due to the well-known
difficulties of the nonperturbative quantum chromodynamics (QCD), is
much more complicated for theoretical analysis and  hence  represents
a leading contributor to uncertainties in   theoretical
predictions. Electroweak corrections are suppressed by the
$Z-$bozon mass as $m_L/M_Z$ and can contribute only at the seventh
significant digit in calculations and are of the same order as
corrections from the diagrams with light by light scattering.  A
comprehensive analysis of the role of different mechanisms in to the
lepton anomaly can be found, e.g. in Ref.~\cite{review-2021}.  The
net result is that with adding together all the  mentioned mechanisms, the
deviation of theoretical calculations from experimental data for
electrons $\Delta a_{e} \equiv
a_{e}^{\mathrm{exp}}-a_{e}^{\mathrm{SM}}  \simeq (-87 \pm 36)\times
10^{-14}$ is as large as $ \sim 2.5  \,$   standard
deviations~$\sigma$, see
Refs.~\cite{Parker:2018vye,Davoudiasl:2018fbb}, whereas, according to
the most recent measurements of the E989 experiment  at
Fermilab~\cite{E989} together with previous measurements of the muon
anomaly of the E821 experiment at BNL~\cite{E821},  the muon deviation
is found to be $ \Delta
a_{\mu}=a_{\mu}^{\mathrm{exp}}-a_{\mu}^{\mathrm{SM}}=(251\pm59)\times
10^{-11}\sim 4.2\, \sigma \,$. It should be stressed that  the
contribution of the HVP mechanism  is usually  estimated either using the
dispersive technique combined with ``$e^+ e^-\to hadrons$'' cross
section data, see Refs.~\cite{Malaescu,Nomura,HJEP,Kubis}, or using  cumbersome and
lengthy lattice QCD calculations, cf., Ref.~\cite{lattice} (for more
detailed discussions, see Ref.~\cite{Jul2022} and  references therein).
If we  rely   on the latest  lattice calculations of the HVP
corrections,   the discrepancy $\Delta a_{\mu}$ becomes  reduced up to
$\sim 1.5\ \sigma$. Nevertheless, it is essential that this result be
confirmed by independent lattice
calculations~\cite{prospects,Kuberski}. Note, that the systematic
theoretical results overestimates of the electron anomaly and underestimates
of the muon anomaly. These circumstances  motivate  future
experimental investigations of  lepton magnetic moments,  viz.,  at
Fermilab~\cite{Grange:2015fou} and J-PARC \cite{Iinuma:2011zz}   and
 renew interest  in improving the accuracy of
theoretical calculations  within  the mentioned mechanisms.

In the present paper, we consider corrections solely from the pure QED
contributions from a subset of Feynman diagrams that allow one to obtain
close analytical expressions for the lowest  order, up to the fourth,
in the fine structure constant $\alpha$. The very first calculations
of the radiative corrections to the electron giromagnetic factor were
performed by J.~S.~Schwinger~\cite{Schwinger1948} who obtained
the electron anomaly to be $\alpha/( 2\pi )\simeq 0.00116$ in
excellent agreement with the  experimental
data, available at that time. Further increase in measurement  accuracy requires more refined
theoretical calculations of the QED contribution, e.g., corrections of
the eighth-, tenth- and higher order w.r.t. electromagnetic coupling
constant. So far,  higher order analysis has been  based mainly on
either   approximate asymptotic expansion of the corresponding Feynman
diagrams  or  more accurate but cumbersome and computer time
consuming  numerical calculations, cf.
Refs.~\cite{Jegerlehner:2017gek,Kinoshita-1990,Laporta:1993ju,Laporta:1993ds,Kinoshita:2005zr,Laporta:2017okg,Aguilar:2008qj,
Kurz:2013exa,Kurz:2016bau,Baikov,Marquard:2017iib} and references
therein.
However, as far as we know, the corresponding higher order exact
analytic expressions, which can serve  as serious tests for both
asymptotic formulas and numerical results, have  not been presented in
the literature. In this context, it is rather appealing to separate
from the full set of diagrams of a given order at least  a subset
that permits one to perform calculations in an explicit analytical form.
This type  subset consists exclusively of diagrams with insertions of the
photon polarization operator with at least three/four closed lepton loops,
the so-called bubble-like diagrams. Certainly,   the explicit expressions
are more attractive since they allow  performing calculations with any
desired accuracy and  also testing the reliability of  asymptotic
expansions and numerical evaluation of the corresponding integrals.

In this paper, we consider this type of  QED Feynman diagrams with insertions of
the photon polarization operator and derive analytical expressions for
the radiative corrections up to the eighth order to anomalous
magnetic moments of leptons. In this sense, the paper can be viewed as
a generalization  of the results previously reported in the literature concerning mainly  muons,
cf. Refs.~\cite{Laporta:1993ju,Laporta:1993ds,Aguilar:2008qj,Friot:2005cu},
to all types of leptons,\ $e, \mu$ and $\tau$ and to the whole region of the mass ratio, $0 <\, m_\ell/m_L\, <\, \infty$, where $m_L$ and $m_\ell$ denote the mass  of the considered lepton $L$ and the mass of the loop  leptons,
respectively.
The considered diagrams refer to any   lepton  and include all
possible combinations of leptons in the polarization operator with maximum three
loops formed by  two different or three identical leptons.

The paper is organized
as follows. In order to facilitate the reading of the paper, in
Section~\ref{Prelim} we briefly recall the main definitions relevant  to
calculations of the lepton anomaly. The general relation between the
anomaly induced by the polarization operator of the virtual photon
with an arbitrary number of closed lepton loops and the anomaly due to the
exchange of a single  but massive  photon is established.
Section~\ref{basic} is dedicated to calculations of the Feynman
bubble-like diagrams of any order and applications of dispersion
relations and the Mellin-Barnes transform to the corresponding
$x-$parametrizations of Feynman integrals. The theoretical approach is
basically the one developed by  E.~de~Rafael and
coauthors~\cite{Friot:2005cu,Aguilar:2008qj} for investigations of the muon anomaly.
 Previously, a similar approach
was applied to calculate analytically  the anomaly of muons up to
the sixth order w.r.t. the electron charge in
Ref.~~\cite{Laporta:1993ju}. Here we generalize the method to any type
of leptons with all possible insertions in the polarization operator
to determine analytically  radiative corrections up to the eighth
and, possibly, higher order. Sections~\ref{Sec:one}-\ref{Sec:three} are entirely
devoted to establishing analytical expressions for   radiative
corrections from diagrams with one, two and three loop insertions,
respectively. The radiative corrections are expressed in terms of the
mass-ratios $r=m_\ell/m_L$ of the internal  to  external masses of the concerned
leptons. Particular attention is paid to the determination, for each type of
diagrams, of the generic analytical function valid in the whole interval
of $r$  which, as shown below,  represents an analytical continuation
of the corresponding correction derived separately for $r<1$ and
$r>1$. Once such functions are determined explicitly, one can perform
numerical calculations for the corrections of a given order with any
desired precision. In Section~\ref{results}, we present a qualitative
numerical analysis of the obtained analytical expressions for the
corrections up to the eighth order. We investigate the dependence of
corrections originating from all possible types of  loop insertions
as a function of the variable~$r$. We argue that in the Feynman diagram
the contribution of the lepton loops decreases with increasing
mass of the internal leptons. The hierarchy of three loop diagrams is
determined (numerically) for each type of the considered leptons. In
Subsection~\ref{Subsec:asympt}, we  compare our results
with the known asymptotical expansions at $r\ll 1$ reported in the
literature and augment them by the asymptotics of $A_2^{(8)}(r)$ for $r\gg 1$, not yet considered hitherto.
Summary and conclusions are collected in Section~\ref{summary}.
Eventually, some useful relations among the relevant special functions
which can facilitate comparisons with previous results obtained by
different authors are presented in   Appendix~\ref{app}.

\section{Lepton anomaly and radiative corrections }\label{Prelim}

In order to investigate the magnetic property of the lepton $L$ (electron, muon or tau-lepton),
one  considers the scattering of the lepton $L$ in an external magnetic field $A_\mu^{(ext)}(q^2)$.
The corresponding   scattering amplitude is stipulated by only two scalar functions $F_1(q^2)$  and $F_2(q^2)$:
\begin{eqnarray} && \hspace*{-5mm}
T(p_1,p_2)=e \bar u(p_2) \Gamma_\mu (p_1,p_2) u(p_1)A^{\mu (ext)}(q^2)= e\bar u(p_2)\left[ \gamma_\mu F_1(q^2) +
i \frac{\sigma_{\mu\nu} q^\nu}{2m_L} F_2(q^2)\right ] u(p_1)A^{\mu (ext)}(q^2),\nonumber \\ &&
  \label{odin1}
\end{eqnarray}
where the initial and final lepton momenta are $p_1$ and $p_2$, respectively,
$q=p_2-p_1$, $A^{(ext)} = A^{(ext)}(0, {\bf A})$ with ${\bf B}={\rm rot}\ {\bf A}$ and the electromagnetic
vertex $\Gamma_\mu$ is
\begin{eqnarray}
\Gamma_\mu =\left[ \gamma_\mu F_1(q^2) +
i \frac{\sigma_{\mu\nu} q^\nu}{2m_L} F_2(q^2)\right ].
\label{gamma1}
\end{eqnarray}
The Gordon identity \ba \bar u(p_2)  \gamma^\mu  u(p_1) =
\frac{1}{2m_L} \bar u(p_2)\left
[(p_1+p_2)^\mu+i\sigma^{\mu\nu}q_\nu\right] u(p_1) \label{dva-Prelim}
\ea
allows one to rewrite the scattering amplitude (\ref{odin1}) as
\ba
T(p_1,p_2)=e\bar u(p_2)\Gamma_\mu  u(p_1)A^{\mu (ext)}(q^2).
\label{tri} \ea with \ba \Gamma_\mu =\left[\frac{(p_1+p_2)^\mu}{2m_L}
F_1(q^2) + i \frac{\sigma_{\mu\nu} q^\nu}{2m_L} \left
(F_1(q^2)+F_2(q^2)\right)\right ]. \label{gamma2}
\ea
Obviously, for
the on shell leptons, Eqs.~(\ref{gamma1}) and (\ref{gamma2}) are
completely equivalent. In the non-relativistic   and static limit of
$q\to 0$, only $\sigma_{ij}$, ($ij  = 1, 2, 3$ ) contributes to
$\Gamma_\mu$   and the second term in ({\ref{gamma2})   reduces to
\ba
-\frac{e}{2m}(1+F_2(0)){{\psi}^+} \boldsigma\cdot{\bf  B}\;{\psi}
\to -g_L\boldmu\,  {\bf  B}, \ea
where $\boldmu=\left(\displaystyle\frac{e}{2m}{{\bf s}} \right) $ is the magnetic moment of an elementary particle with
spin   ${\bf  s} =\boldsigma/2$  and  the gyromagnetic ratio $g_L=  2(1 +F_2(0))$ is the
measure of the magnetic anomaly of the lepton. In practice, it is more
convenient  to consider another quantity,
\ba
a_L=\displaystyle\frac{g_L-2}{2} = F_2(0). \label{gminusdva}
\ea

It is therefore clear that to find the lepton magnetic anomaly $a_L$
one should express the calculated (fully dressed) electromagnetic
vertex $\Gamma_\mu$ in the form of Eq.  (\ref{gamma1}) or
(\ref{gamma2}),  take the limit $q\to 0$ and find   the coefficient in
front of the operator $i {\sigma_{\mu\nu} q^\nu}/{2m_L}$. An
alternative and rather elegant method  for determining   $a_L$ consists in
 employing  a  properly defined   projection operator ${\cal P}_\mu$
that,  acting on $\Gamma_\mu$, separates   $a_L$  and, consequently, the
gyromagnetic factor $g_L$. For instance, one can define
the following projection operator~\cite{proj}:
\ba
&&{\cal P}_{\mu} =  \frac{1}{Q^2}(\gamma_\mu \hat {q}  +p_\mu \hat{q}/m_L  - q_\mu )(\hat{p} + m_L)+
\frac13\gamma_\mu -\frac{p_\mu}{m_L}-\frac43 p_\mu \frac{\hat{p}}{m_L^2},\label{proj1}
\ea
where $p=(p_1+p_2)/2$ and $q=(p_2-p_1)$ with $(p\cdot q)=0$.
Then
\ba
a_L=\lim_{q\to 0}\left[ \frac14 Tr \left ({\cal P}^\mu \Gamma_\mu\right)\right ]. \label{trac}
\ea

Theoretically, to calculate the anomalous magnetic moment of a lepton
up to the desired order, one should consider radiative corrections
to the electromagnetic vertex $\Gamma_\mu (p_1,p_2)$ and fold the
obtained expression with the projection operator Eq.~(\ref{proj1}) and
use Eq.~(\ref{trac}). In the present work, we
consider the radiative corrections due to Feynman diagrams
  with insertions in to the virtual photon propagator of the vacuum polarization operator. The
corresponding  diagram is depicted in Fig.~\ref{f1}, left panel, where
the photon propagator is
\begin{figure}
%\phantom{}\vspace{0.8cm}%
%\includegraphics[width=0.5\textwidth]{GammaFeyn.eps}
\includegraphics[width=0.40\textwidth]{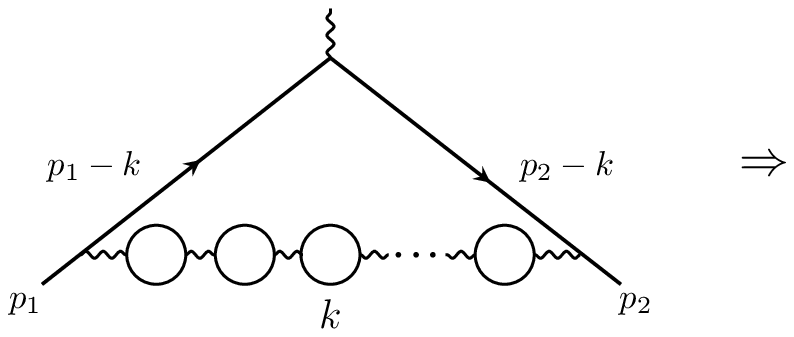}\hspace*{1cm}
\includegraphics[width=0.35\textwidth]{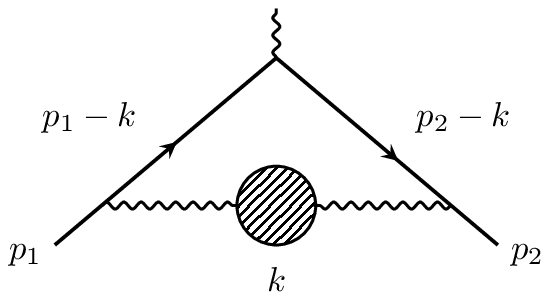}
%\vspace*{0.1cm}
\caption{Radiative corrections to the lepton electromagnetic vertex due to  vacuum polarization
insertions, left panel, and the equivalent diagram with vacuum polarization of a massive photon, right panel. }
\label{f1}
\end{figure}
\ba &&
D_{\alpha\beta}(k^2) =  -ig_{\alpha\beta}\frac{1}{k^2}
\frac{1}{1+\Pi(k^2)}=\nonumber\\ &&
-ig_{\alpha\beta}\frac{1}{k^2}\left[
1-\left( \Pi(k^2)+\Pi^2(k^2)-\Pi^3(k^2)+\cdots\right)\right]\equiv
-ig_{\alpha\beta}\frac{1}{k^2}\left[
1- \widetilde\Pi(k^2)\right],
\label{prop}
\ea
where $\widetilde\Pi(k^2)$ is the full polarization operator of the virtual photon. Explicitly, the sought
vertex function $\Gamma_\mu(p_1,p_2)$ is
\ba
\Gamma_\mu(p_1,p_2) = -ie\frac{e^2}{(2\pi)^4}\int d^4k
\gamma_\alpha\frac{(\hat {p_2} -\hat{k} +m_L)\gamma_\mu (\hat {p_1} -\hat{k} +m_L)}
{(k^2-2p_2k)(k^2-2p_1k)}\gamma_\alpha\frac{ \widetilde\Pi(k^2) }{k^2}.
\ea
The next step is to apply the dispersion relations to the   operator $  -\widetilde\Pi(k^2)/k^2$,
\ba &&
\Gamma_\mu(p_1,p_2) = -ie\frac{e^2}{(2\pi)^4}\int \frac{dt}{t}
\frac1\pi\frac{\Imm \widetilde\Pi(t)}{k^2-t}\int d^4k
\gamma_\alpha\frac{(\hat {p_2} -\hat{k} +m_L)\gamma_\mu (\hat {p_1} -\hat{k} +m_L)}
{(k^2-2p_2k)(k^2-2p_1k)}=\nonumber\\ [0.1cm]
&&
\frac{1}{\pi} \int \frac{dt}{t}
 {\Imm} \widetilde\Pi(t)\left[ -ie\frac{e^2}{(2\pi)^4}\int d^4k
\gamma_\alpha\frac{(\hat {p_2} -\hat{k} +m_L)\gamma_\mu (\hat {p_1}
-\hat{k} +m_L)} {(k^2-2p_2k)(k^2-2p_1k)}\gamma_\alpha
\frac{1}{k^2-t}\right],
\ea
 where the expression in  square brackets
is nothing but the second order correction to the vertex
$\Gamma_\mu^{(2)}(p_1,p_2,t)$  from diagrams with the  exchange of one
massive photon with the mass $m_\gamma^2=t$, see right panel in
Fig.~\ref{f1}. Then
 \ba
 && \Gamma_\mu(p_1,p_2) = \frac1\pi\int \frac{dt}{t}
 \Imm \widetilde\Pi(t) \Gamma_\mu^{(2)}(p_1,p_2,t).\label{massive}
\ea
In such a way, the corrections to the  electromagnetic vertex
$\Gamma_\mu$ of  an arbitrary order can be expressed via the
polarization operator $\tilde\Pi(t)$ and the second order
electromagnetic vertex of a massive photon. This implies   that  the
anomalous magnetic moment $a_L$ is also determined by the second order
anomalous magnetic moment $a_L(t)$ of a massive ($ m_\gamma^2 =  t$)
photon  folded with the polarization operator $\Imm\, \tilde\Pi(t)$. The
former is known explicitly~\cite{berestetski,brodski}  and one can
write
\ba a_L&&=\frac1\pi\int \frac{dt}{t}
 \Imm \widetilde\Pi(t) F_2^{(2)}(t)
 =\frac1\pi\int \frac{dt}{t}
 \Imm \widetilde\Pi(k^2)\frac\alpha\pi\int dx\frac{x^2 (1-x)}{x^2+(1-x)t/m_L^2}=\nonumber \\ &&
 -\frac\alpha\pi\int dx (1-x) \widetilde\Pi(q_{eff}^2),\label{double}
\ea
where $\alpha=e^2/4\pi$ is the fine structure constant and the effective momentum $q_{eff}$ is defined as
\ba
  q_{eff}^2=-\frac{x^2}{1-x}m_L^2 . \label{qeff}
\ea
Note the Euclidean nature of $q_{eff}^2 <0$  and   that $$\widetilde\Pi(k^2)  =\Pi(k^2)-\Pi^2(k^2)+\Pi^3(k^2)-\cdots \,  .$$

\section{Basic formalism: Mellin-Barnes integral representations}\label{basic}

Although Eq.~(\ref{double}) can be considered as the final expression
for numerical calculations of $a_L$ up to the desired order, in this
paper we focus on  revealing  the  prerequisites for explicit
analytical expressions for $a_L$.  The general method of analytical
consideration of $a_L$ was presented in
Refs.~\cite{Friot:2005cu,Aguilar:2008qj,Friot:2011ic,Rafael-HVP} where
 the sixth-order corrections to the muon anomaly were
examined in some detail. Below we use the  method for any
type of lepton and apply it to find analytically the corresponding corrections up to
the eighth order. As in Ref.~\cite{Aguilar:2008qj}, we consider
diagrams with lepton loops of the polarization operator consisting of two different
leptons, i.e. diagrams for which
$\Pi(k^2)= \left ( \Pi^{(l_1)}(k^2) + \Pi^{(l_2)}(k^2)\right) $.
Moreover, one of the internal leptons is chosen  to be  of the same kind
as the external one. The case of identical internal loops is included as well.
The case of diagrams depending on three masses,
$m_e$, $m_\mu$ and $m_\tau$, or, equivalently,  on two mass ratios
$r_1$ and $r_2$, are more complicated for analytical calculations.  The
first exact expressions for this case were obtained only for the
sixth order diagrams~\cite{Ananthanarayan:2020acj}, i.e. for diagrams
with two different internal loops  other than   the external lepton. An
analytical analysis of such corrections will be
represented elsewhere.

 Generally,  the polarization operator for Feynman diagrams with
 insertions of $n=p+j$ closed loops, where $p$  and $j$ denote the
 number of leptons  loops of   $\ell_1$ and $\ell_2$ kinds, can be presented
 as
  \ba &&
  \Pi^{ n }(k^2)=  \sum\limits_{p=0}^n (-1)^{n+1} C_n^p\left[
  \Pi^{(\ell_1)}(k^2)\right ]^p \left[ \Pi^{(\ell_2)}(k^2)\right ]^{j=n-p}\equiv
    \sum\limits_{p=0}^n F_{(p,j)}\left[ \Pi^{(\ell_1)}(k^2)\right ]^p \left[ \Pi^{(\ell_2)}(k^2)\right ]^{j=n-p},
 \nonumber \\
\label{pandj}
\ea
\vspace*{-0.2cm}\noindent
where $C_n^p$ are the combinatorial coefficients and the quantity $
F_{(p,j)} =(-1)^{p+j+1} C_{p+j}^p $  has been introduced as  to
reconcile our formulae with the commonly adopted
 notation, cf. Ref.~\cite{Aguilar:2008qj}.
Note that   $p$ and $j$ enter symmetrically  in the polarization operator $\Pi^{ n }$, i.e. one can interchange
$\ell_1\leftrightarrow   \ell_2$ in  Eq.~({\ref{pandj}).   In what follows  we choose the leptons   $\ell_1$ to be
of the same kind as the external one. Consequently,
for the sake of  brevity,  below we release the labels $\ell_1$, $\ell_2$  for the loop leptons and  use merely
the notation $\ell_1=L$, and $\ell_2=\ell$.
Then the contribution to the anomalous magnetic moment from diagrams with
$p$ leptons of type $L$   and $j$ leptons of type $\ell$   (c.f. Eqs.~(\ref{double}) and (\ref{pandj})) acquires the form
  \ba &&
 a_L^{(p,j)}=
\frac\alpha\pi F_{(p,j)} \int_0^1 dx \; x^2 (1-x) \left[ \Pi^{(L)}(q_{eff}^2)\right ]^p \frac1\pi
\int_0^\infty\frac{dt}{t} \frac{ \Imm
\left[ \Pi^{(\ell)}(t)\right ]^j}{ x^2 +(1-x) t/m_L^2}=\nonumber\\ &&
 F_{(p,j)}\frac\alpha\pi \int\limits_0^\infty\frac{dt}{t} \int\limits_0^1 dx \frac{x^2 (1-x)}{x^2 +(1-x) t/m_L^2}
\left[ \Pi^{(L)} \left(-\frac{x^2 }{1-x)}m_L^2  \right)\right ]^p
\frac1\pi \; \Imm \left[ \Pi^{(\ell)}(t)\right ]^j,\label{secnd}
  \ea
where, as mentioned, the external lepton is labeled as $L$.
This is the well known expression obtained by E.~de Rafael and coauthors~\cite{Aguilar:2008qj} for
the muon anomaly governed  by
the polarization operator  with $n$ closed loops
of two different leptons. Usually one introduces a new notation
\ba
\rho_j\left(\frac{4m_{\ell}^2}{t}\right)=\frac1\pi \Imm \left [\Pi^{(\ell)}(t)\right]^j \, ,
\label{rho}
\ea
which is inspired by the fact that  actually the polarization operator  depends rather on the combination
$4m^2/t$ than  solely  on $t$.

Equation (\ref{secnd}) is the main expression  suitable
for further   analytical calculations of $a_L$  which will allow one to
obtain $a_L$ with the  precision as high  as permitted by the
knowledge of the fundamental constants entering into $a_L$, viz., the fine
structure constant and lepton masses $m_e$, $m_\mu$ and $m_\tau$. To
this end, let us consider the known Mellin-Barnes integral
representation for propagator-like functions of a massive scalar
particle, cf.~Ref.~\cite{boos}. One starts with the known integral
representation of the Euler beta-function
   \begin{equation}
   B(s,\xi)=\int\limits_0^\infty dx \; \frac{x^{s-1}}{(1+x)^{s+\xi}} \, , \label{odin}
   \end{equation}
which, for  a particular choice  of the variables $s$ and $\xi$, namely for  $\xi = \beta-s$, reads as
   \begin{equation}
   B(s,\beta-s)=\int\limits_0^\infty dx \; \frac{x^{s-1}}{(1+x)^{\beta}}=
   \int\limits_0^\infty dx \;  x^{s-1} F(x,\beta)  ,\label{dva-beta}
   \end{equation}
i.e., it can be considered as the Mellin transform of the   function
   \ba
   F(x,\beta)=\displaystyle\frac{1}{(1+x)^\beta}. \label{ee}
   \ea
Then, the inverse Mellin transform  of $F(x,\beta)$ is
\begin{equation}
   F(x,\beta)=\displaystyle\frac{1}{(1+x)^\beta}= \frac{1}{2\pi i}\int\limits_{c-i\infty}^{c+i\infty}  ds \ x^{-s} \  B(s,\beta-s) =
 \frac{1}{2\pi i}\int\limits_{c-i\infty}^{c+i\infty}  ds \  x^{-s}\
 \frac{\Gamma(s)\ \Gamma(\beta-s)}{\Gamma(\beta)} \, ,
 \label{tri-MB}
 \end{equation}
 where $0 <  c < \beta$.  Equation (\ref{tri-MB}) is known as the Mellin-Barnes representation for propagator functions.
Coming  back to Eq.~(\ref{secnd}), let us apply it to the integrand in integration over $x$
\ba
\frac{x^2 (1-x)}{x^2 +(1-x) t/m_L^2} = \frac{1}{2\pi i}\int\limits_{c-i\infty}^{c+i\infty}
ds \;  \left( \frac{4 m_{\ell}^2}{t}\right)^s\left( \frac{4m_{\ell}^2}{m_L^2}\right)^{-s}
x^{2s}(1-x)^{1-s} \; \Gamma(s) \Gamma(1-s), \label{m-b}
\ea
where the quantity $4m_\ell^2$ has been introduced for further convenience.
With this representation, Eq.~(\ref{secnd}) reads as
\ba
a_L(p,j)&&=\frac{\alpha}{\pi}\frac{1}{2\pi i}F_{(p,j)}
\int\limits_{c-i\infty}^{c+i\infty}  ds \; \left (\frac{4m_{\ell}^2}{m_L^2}\right)^{-s}
\Gamma(s)\Gamma(1-s)\int_0^1 dx \; x^{2s} (1-x)^{1-s}
\times\nonumber  \\ &&
 \left[ \Pi^{(L} \left(-\frac{x^2 }{1-x)}m_L^2  \right)\right ]^p
\int_0^\infty\frac{dt}{t}\left( \frac{4m_{\ell}^2}{t}\right)^s\rho_j\left(\frac{4m_{\ell}^2}{t}\right)
 ,
\label{fin}
\ea
where $0< c < 1$.
It can be seen that the Mellin-Barnes transform made it possible to  present the contribution
to the lepton anomaly   from different kinds of lepton loops in the following factorized form of two Mellin momenta
 \ba &&
a_L(p,j)=\frac{\alpha}{\pi}
\frac{1}{2\pi i}F_{(p,j)}
\int\limits_{c-i\infty}^{c+i\infty} ds \;
 \left( \frac{4m_{\ell}^2}{m_L^2}\right)^{-s}
\Gamma(s)\Gamma(1-s)\; \left(\frac{\alpha}{\pi}\right)^{p}\Omega_p(s) \left(\frac{\alpha}{\pi}\right)^{j}R_j(s),
\label{fin1}
\ea
where
\ba &&
\left(\frac{\alpha}{\pi}\right)^p\Omega_p(s)=
  \int_0^1 dx \; x^{2s} (1-x)^{1-s}
  \left[ \Pi^{(L)} \left( -\frac{x^2 }{1-x}m_L^2  \right)
  \right ]^p \label{Omp}
  \\ &&
\left(\frac{\alpha}{\pi}\right)^j R_j(s)=
\int_0^\infty \frac{dt}{t}\left( \frac{4m_{\ell}^2}{t}\right)^s\rho_j\left(\frac{4m_{\ell}^2}{t}\right).
\label{Rj}
\ea

It is  seen from (\ref{fin1})-(\ref{Rj}), that to compute   $a_L$ up to
the    $2(p+j+1)$ order  it suffices to calculate separately the
polarization operators  $\Pi^{(L)}(q^2)$ and  $\Imm\,
\Pi^{(\ell)}(q^2)$. Recall that  the generic variable for the
$q^2$-dependence of  the operator $\Pi(q^2)$ is the dimensionless
combination $y=\frac{4m^2}{q^2}$, i.e. $\Pi^{(L)}(q^2)=\Pi^{(L)}(y)$.
It implies that the quantity $\Omega_p(s)$, for which $y=-4(1-x)/x^2$,
does not depend on  the lepton mass $m_L$. Furthermore, the term
relevant to the second lepton, $R_j(s)$ \ba
 \left(\frac{\alpha}{\pi}\right)^j   R_j(s)=
\int_0^\infty\frac{dt}{t}\left( \frac{4m_{\ell}^2}{t}\right)^s\rho_j\left(\frac{4m_{\ell}^2}{t}\right)=
\int_0^\infty d \xi \; \xi^{s-1} \rho_j(\xi)
\label{Rj1}
\ea
is  mass independent as well. Hence, the mass-dependence  in
Eq.~(\ref{fin1}) enters only via the ratio $r=\frac{m_{\ell}}{m_L}.$
Therefore, it is commonly adopted to classify the contribution to the
anomalous magnetic moment $a_L$  by this ratio
\begin{equation}\label{aL}
a_{L} =  ~A_{1}\left( \frac{m_L}{m_L}\right ) +
~A_{2}\left( \frac{m_{\ell}}{m_L}\right )  + A_{3}\left(\frac{m_{\ell}}{m_L},\frac{m_{\ell_2}}{m_L}\right ),
\end{equation}
where $A_1$ corresponds to   diagrams  for which all the  internal
loops are formed by the same leptons as  the external one. It also
includes the case of diagrams without lepton loops, i.e. the diagrams
of the second order with respect to the electromagnetic coupling $e$
with exchange of only one virtual photon. Clearly, the coefficients
$A_1$ are  universal for all kinds of leptons. The coefficients $A_2$
correspond to  diagrams with either two different types of internal loops   with  at least one
  coinciding with  the external lepton
$ L$ and the other  $\ell\neq L$, or with   loops all formed by  identical leptons
of type $\ell$,
$\ell\neq L$.
Eventually, $A_3$ correspond to diagrams with loops formed
by maximum three kinds of leptons with at least two non identical
 and  different from the external lepton, $l\neq l_2 \neq L$.
Consequently, in dependence of the number of   insertions of the
polarization operators,   each   coefficient $A_i$ can be classified
by the corresponding number of bubble-like loops $n$ ($n=0,1...$) or,
equivalently, by the  $(n+1)$-th    power of $\alpha$ and computed order by order.
\ba
&& A_1(m_L/m_L)= A_1^{(2)}\left(\frac\alpha\pi \right)^1 + A_1^{(4)}\left(\frac\alpha\pi \right)^2 +
A_1^{(6)}\left(\frac\alpha\pi \right)^3 + \cdots , \label{A1}\\[1mm]
&& A_{2}\left( r={m_{\ell}}/{m_L} \right) = A_{2}^{(4)}(r)\left(\frac\alpha\pi \right)^2
+ A_{2}^{(6)}(r)\left(\frac\alpha\pi \right)^3 +
A_{2}^{(8)}(r)\left(\frac\alpha\pi \right)^4 + \cdots,\label{A23} \\[1mm]
&& A_3\left(r_1,r_2\right)=
A_3^{(6)}(r_1,r_2)\left(\frac\alpha\pi \right)^3 + A_3^{(8)}(r_1,r_2)\left(\frac\alpha\pi \right)^4 +
A_3^{(10)}(r_1,r_2)\left(\frac\alpha\pi \right)^5 + \cdots ,~~~ \label{A4}
\ea
where $r_1= {m_{\ell_1}}/{M_L},~r_2={m_{\ell_2}}/{M_L}  $ and the superscripts of the coefficient in the r.h.s.  indicate the
corresponding order of the radiative corrections, while the powers of
$\alpha^{n+1}$  correspond to the  number $n$ of loops in the
bubble-like Feynman diagrams. N.B:   The diagram without loops
($n=0$), as mentioned,   was calculated explicitly by
Schwinger~\cite{Schwinger1948} and found to be equal to $
\alpha/2\pi   $, i.e.,  in our notation $ A_1^{(2)}=1/2 $.
Moreover, it can be seen form   Eqs.~(\ref{secnd}), (\ref{Omp}) and
(\ref{A1})  that each of the coefficients $A_1$ in (\ref{A1}) is
uniquely  determined by the corresponding value of $\Omega_p(s=0)$.
For instance,  for $n=0$ one has $\Omega_0(s=0)=A_1^{(2)}=\frac12$
  which is exactly the normalization of (\ref{aL}) to the
Schwinger term $ {\alpha}/{2\pi}$. The  mass-independent coefficients
$A_1$ are less complicated and have been thoroughly investigated
previously;  their general explicit analytical expressions
 can be found
 in Refs.~\cite{Samuel-n-bubble,Jegerlehner:2017gek, Petermann:1957hs,Sommerfield:1957zz,Laporta:1996mq,Laporta:2018ngl}.

In the present paper, we consider the radiative corrections up to the eighth order, i.e. the
coefficients $A_{1,2}^{(4)-(8)}(r)$ in Eqs.~(\ref{A1})-(\ref{A23}). They include diagrams with one, two and three lepton loops.
Since calculations of the coefficients $A_2(r)$ presuppose also
explicit calculations of $\Omega_p(s)$, the coefficients $A_1$ for
which $r=1$ can be easily inferred from  $\Omega_p(s\to 0)$ or from
$A_2(r\to 1)$.

\section{One lepton loop corrections}\label{Sec:one}

Here  we present in some detail the derivation of the explicit
expression for the one loop $l\neq L$ corrections. The corresponding diagram is depicted in
Fig.~\ref{Fig-2-loop}, left panel.

%%%%%%%%%%  FIGURE 2-loop--  %%%%%%%%%%%%%%%%%%%%
\begin{figure}[!hbt]
\vspace*{4mm}
\includegraphics[width=0.27\textwidth]{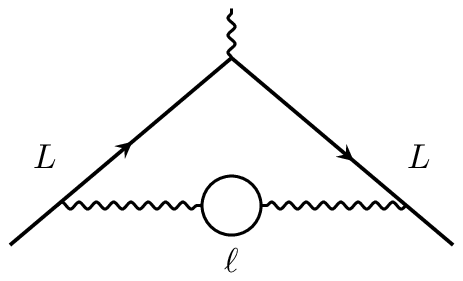} \hspace*{5mm}
\includegraphics[width=0.27\textwidth]{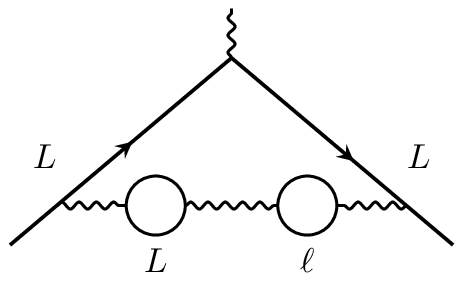} \hspace*{5mm}
\includegraphics[width=0.27\textwidth]{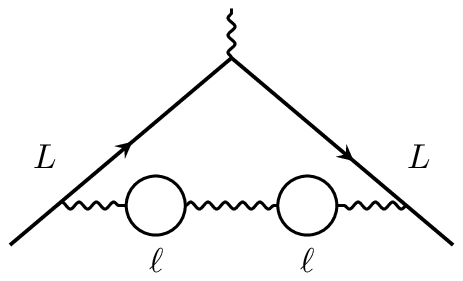}
%\vspace*{-2mm}
\caption{One and two loop corrections to the lepton ($L$) anomalous magnetic moment of the fourth, left panel,
and sixth,  central and right panels, order determined by the  coefficients
 $A^{(4,6)}_2(r=m_l/m_L)$ in Eq.~(\ref{A23}). Left panel: one loop diagram with the internal
 lepton $(\ell)$  different from $L$   for which
 $ \Omega_p(s)=\Omega_0(s)$  and  $R_j(s)=R_1(s)$.
 Central panel:  diagrams with two different  $(Ll)$ leptons for which
 $ \Omega_p(s)=\Omega_1(s)$  and  $R_j(s)=R_1(s)$.
 Right panel:   diagrams with two identical $(ll)$ leptons ($l\neq L$) for which
$\Omega_p(s)=\Omega_0(s)$ and $R_j(s)=R_2(s)$, cf. Eqs.~(\ref{Omp}) and (\ref{Rj}).} \label{Fig-2-loop}
\end{figure}
 In this case   in
Eq.~(\ref{fin1})  $p=0$, $j=1$ and the corresponding anomalous
magnetic moment is
 \ba &&
a_L(0,1)=\frac{\alpha}{\pi}
\frac{1}{2\pi i}
\int\limits_{c-i\infty}^{c+i\infty}  ds
 \left(  4r^2 \right)^{-s}
\Gamma(s)\Gamma(1-s) \; \Omega_0(s)\; \left(\frac{\alpha}{\pi}\right) R_1(s),
\label{fin2}
\ea
where, according to Eqs.~(\ref{Omp}) and (\ref{Rj})
\ba &&
\Omega_0(s)=\int_0^1 dx \; x^{2s} (1-x)^{1-s}=
\frac{\Gamma(2-s) \Gamma(1+2s)}{\Gamma(3+s)},
   \label{Omp0}
    \\ [0.1cm] &&
\left(\frac{\alpha}{\pi}\right)  R_1(s)=
\int_0^\infty\frac{dt}{t}\left( \frac{4m_l^2}{t}\right)^s\rho_1\left(\frac{4m_l^2}{t}\right)
\label{Rj11}.
\ea

The polarization operator of photons  $\Pi^{(\ell)}(q^2)$ in QED  is known
explicitly and reported in a series of publications, see e.g.
Refs.~\cite{Jegerlehner:2017gek,Aguilar:2008qj}.
With   $\delta=\sqrt{1- {4m_\ell^2}/{q^2}}$, it reads as
\ba &&
\Ree \; \Pi^{(\ell)}(q^2) = \left(\frac{\alpha}{\pi}\right )
\left[\frac89 - \frac{\delta^2}{3}+\delta \left(\frac12 -\frac{\delta^2}{6}\right)\ln\frac{|1-\delta|}{1+\delta}\right],
\label{reP1} \\[0.2cm] &&
\frac1\pi \Imm  \;  \Pi^{(\ell)}(q^2) = \left(\frac{\alpha}{\pi}\right ) \delta
\left(\frac12 -\frac16 \delta^2\right)\theta\left (q^2-4m_\ell^2\right).
\label{imP1}
\ea

Evidently,  since the Euclidean nature of the effective momentum $
q_{eff}$, c.f.  Eq.~(\ref{qeff}), and due to the presence of
$\theta\left (q^2-4m_\ell^2\right)$ in (\ref{imP1}),   the
polarization operator $\Pi(q_{eff}^2)$ in~(\ref{Omp}) is pure real and
 can be written as~\cite{Lautrup77},
\ba &&
\Pi^{(L)}(q_{eff}^2)=\frac\alpha\pi f(x), \,\, {\rm with} %\nonumber \\ &&
\,\, f(x)=\frac59 +\frac{4}{3x} - \frac{4}{3x^2}+
\left (-\frac13 +\frac{2}{x^2}-\frac{4}{3x^3}\right ) \ln(1-x).
\label{lashk}
\ea
Consequently $\Omega_p(s)$ in~(\ref{Omp}) is also pure real. Yet,  the function
$\theta\left (q^2-4m_\ell^2\right)$ in (\ref{imP1}) restricts
integrations over $t$ in Eq.~(\ref{Rj1}) to the interval  $[4m_\ell^2
\ , \  \infty]$, which is converted in to $\xi\in [0 \ , \  1]$  in
Eqs.~(\ref{Rj1}) and (\ref{Rj11}).
Then,  direct calculation of the integral (\ref{Rj11})  provides
 \ba
R_1(s)=\frac{\sqrt{\pi}}{~4}\frac1s\frac{~\Gamma(2+s)}{\Gamma(5/2+s)}.
\label{R1s} \ea
 Inserting Eqs.~(\ref{Omp0}) and (\ref{R1s}) into
Eq.~(\ref{fin2}), the coefficient $A_2^{(4)}(r)$ reads as
\ba
A_2^{(4)}(r)= \frac{1}{2\pi i} \int\limits_{c-i\infty}^{c+i\infty}  ds
\;
 \left(  4r^2 \right)^{-s}
\Gamma(s)\Gamma(1-s)  \frac{\Gamma(2-s) \Gamma(1+2s)}{\Gamma(3+s)}
\frac{\sqrt{\pi}}{4}\frac1s\frac{\Gamma(2+s)}{\Gamma(5/2+s)}.
\label{fin23}
\ea

It is worth emphasizing that in Eq~(\ref{fin2}) the ratio $r$ acts as
an external parameter so that the integral, if exists, defines
$A_2^{(4)}(r)$ as an analytical  function  of   $r$ in the whole
interval of $r\in (0,\infty)$. Its explicit
expression can be found directly by   integrating   over $s$ in
~(\ref{fin23}), which can be done, e.g. by the Cauchy  residue
theorem. Depending on the value of $r$, one can close the
integration contour to the left $r<1$    or to the right $r>1$
semiplane, respectively, where the integrand (\ref{fin23}) exhibits explicitly  pole-like
singularities with known residues.  The  Cauchy residue theorem
yields

\vskip -0.01cm \noindent
\ba  \label{4thorderLt1}
&& A_2^{(4)}(r<1)=
\frac12\,r\, \left( 5\,{r}^{2}-1 \right)  \left[ {\rm Li_2} \left(r \right) -{\rm Li_2} ( -r  )  \right]
-\,r^4{\rm Li_2}(r^2)\
 +2\,{r}^{4}   \ln^2 ( r )
 + \\ &&
\left\{ \frac12\,r\, \left( 5\,{r}^{2}-1 \right)  \left[ \phantom{\!\!\!\!\!  \frac11}\ln  \left( 1-
r \right) -\ln  \left( 1+r \right)  \right] -\frac13-2\,{r}^{4}\ln
 \left( 1-{r}^{2}\right) +3\,{r}^{2}  \right\} \ln  \left( r \right) -\nonumber \\ &&
~ {\frac{25}{36}}+\frac14\,{\pi}^{2}r-\frac54\,{\pi}^{2}{r}^{3}+4\,{r}^{2}+\frac13\,
{r}^{4}{\pi}^{2}  \, ,\nonumber
\ea
\ba \label{4thorderGt1}
&&  ~~~~ A_2^{(4)}(r>1)=
\frac12\,r\, \left( -5\,{r}^{2}+1 \right)  \left[ {\rm Li_2} \left( 1/r
  \right) -{\rm Li_2} \left( -1/r \right)  \right] -
\frac{25}{36} + r ^{4}{\rm Li_2} \left(1/r^{2} \right) +   \\ &&
~ ~~~\left\{ \frac12\,r\, \left( -5\,r^{2}+1 \right) \left[ \phantom{\!\!\!\!\!  \frac11}\ln (r+1)-\ln( r-1)\right]
 -\frac13+3\,{r}^{2}-2\,{r}^{4}\ln  \left( 1-1/r^{2}
 \right)  \right\} \ln  \left( r \right) +4\,{r}^{2}, \nonumber
 \ea
where ${\rm Li_2}$ denotes the dilogarithm function. As seen,
Eqs.~(\ref{4thorderLt1}) and (\ref{4thorderGt1}) are quite different.
However, using the known properties of complex logarithms and
dilogarithms (see Appendix~\ref{app}), one can show that both equations
are identical in the whole region of the parameter $r \in  (0,\infty)$.
 It means that in practice one can use, for numerical
calculations, any expression (\ref{4thorderLt1}) or
(\ref{4thorderGt1}) regardless of the value of $r$. Nevertheless, one
shall do it with some caution, noting that the logarithms in the square
brackets in (\ref{4thorderLt1}) and   (\ref{4thorderGt1})     should
not  be converted into one logarithm of the ratios of their arguments.
This is because when using (\ref{4thorderLt1}) at  $r>1$, the
complex logarithms $\ln\left[(1+r)/(1-r)\right]$ and $\ln(1+r) -
\ln(1-r)$ differ   by a factor of $2\pi i$.
It also can be shown that using the known relations among the special
functions (see Appendix~\ref{app}) the analytical results for $
A_2^{(4)}(r)$ reported in Ref.~\cite{Friot:2011ic} and  ours,
Eqs.~(\ref{4thorderLt1})-(\ref{4thorderGt1}), are identical. The
analytical continuation of Eqs.~(\ref{4thorderLt1}) and
(\ref{4thorderGt1}) can be written as one function valid in the whole
region~$r\in (0,\infty)$:
 \ba &&
 A_2^{(4),(l)}(r)=   -\frac13 \; \ln(r) {-\frac{25}{36}}
+4\,{r}^{2} +3\;r^2 \ln(r)  + \frac{1}{2} \,r\, \left( 1-5\,{r}^{2}
\right) \bigg[\; {\rm Li_2} \left( \frac{1-r}{1+r} \right)-
\label{4thCommon} \\ [0.12cm] &&
 {\rm Li_2}\left( - \frac{1-r}{1+r} \right) +\frac{1}{4}\pi^2 \;\bigg]
 +
2\left[-{\rm Li_2} \left( \frac{1-r}{1+r} \right)+ {\rm Li_2} \left( -
\frac{1-r}{1+r} \right) +\frac{1}{12}\pi^2 \;- \; 2\; {\rm Li_2}
\left(1-\frac{1}{r}  \right)\right]r^4 \,.\nonumber
 \ea
An analogous expression for the coefficient $A_2^{(4)}(r)$,  in a form slightly different
 from Eq.~(\ref{4thCommon}) form,  was also reported in Ref.~\cite{eidelman}.
%VARIANT 1

Furthermore, we complementary checked that Eqs.~(\ref{4thorderLt1}) and (\ref{4thorderGt1}), as well
as Eq.~(\ref{4thCommon}), reproduce the
well-known asymptotic expansions \footnote{Notice that  the leading
asymptotic terms for $A_2^{(4)}(r)$ werte firstly reported in
Refs.~\cite{suura,Lautrup:1969fr}.}
 \begin{numcases}
{ A_2^{(4),(\ell)}(r)=}   -\frac{1}{3}\; \ln (r)
-\frac{25}{36}+\frac{1}{4}\pi^2 r +\big[ 3+4\ln (r) \big] r^2
-\frac{5}{4} \pi^2 r^3 +  O(r^4); \quad  r\ll 1
\label{1as-loop<1}
\\[0.1cm]
 \frac{1}{45 \; r^2} +
\left[\frac{9}{19600} -\frac{1}{70}\ln (r) \right] \frac{1}{r^4} +O
\left(\frac{1}{r^6}\right);  \quad r\gg 1.
\label{1as-loop>1}
 \end{numcases}
and also the limit
\ba \lim_{r\to 1} A_2^{(4)}(r) = A_1^{(4)} =
\frac{119}{36}-\frac{\pi^2}{3} \, , \label{limA1} \ea which exactly
coincides with the known result obtained before, see e.g.,
Ref.~\cite{Li:1992xf,Czarnecki:1998rc}.

\section{Two lepton loops}\label{Sec:two}

In this section we present calculations of corrections
 due to two  loop  diagrams, as depicted in
Fig.~\ref{Fig-2-loop}, central and right panels, according to  which one  has  two kinds of
contributions.

\subsection{Two identical leptons $(ll)$, $l\neq L$: Fig.~\ref{Fig-2-loop}, right panel}

As follows from expression (\ref{fin1}), in this case $p=0$, $j=2$. Hence the
anomaly reads as
 \ba &&
a_L(0,2)=-\frac{\alpha}{\pi}
\frac{1}{2\pi i}
\int\limits_{c-i\infty}^{c+i\infty} \ ds
 \left(  4r^2 \right)^{-s}
\Gamma(s)\Gamma(1-s) \; \Omega_0(s) \; \left(\frac{\alpha}{\pi}\right)^{2} R_2(s), \label{fin3} \ea
\noindent where $\Omega_0(s)$ is defined by Eq.~(\ref{Omp0}) and
$R_2(s)$  is given  by Eq.~(\ref{Rj}) with
 \ba \rho_2\left(
\frac{4m_l^2}{t}\right) = \frac{1}{\pi}\left[ 2 \Ree \Pi^{(l)}(t)
\Imm \Pi^{(\ell)}(t) \right]. \label{rho2} \ea

Inserting (\ref{rho2}), (\ref{reP1}) and (\ref{imP1}) in to  Eq.~(\ref{Rj}) we get
\ba
R_2(s)=\frac{\sqrt{\pi}}{9} \; \frac{( s-1)( 6+13s+4s^2)} {s^2
(2+s)(3+s)} \; \frac{\Gamma\left(1+s\right)}{\Gamma\left(\frac{3}{2}+s
 \right)}\,.
 \label{R2}
\ea
Applying  in Eqs.~(\ref{Omp0}) and (\ref{R2})  the known
properties of the Euler gamma-functions, e.g. the Euler reflection
and the Legendre duplication formulae, the sixth order contribution
to $a_L$, Eq.~(\ref{fin3}), due to diagrams of the type $(l l)$
 reads as
\begin{equation}
 A_{2}^{(6),(l l)}(r) = \frac{ 1}{2\pi
i}\int\limits_{c-i\infty}^{c+i\infty}  ds\; r^{-2s} \left[
\frac{2(1-s)^2(6+13s+4s^2)}{9s (2s+1) (1+s)(2+s)^2(3+s)}
\right]\frac{\pi^2}{\sin^2(\pi s)}
 \, ,\label{A26}
 \end{equation}
which   manifests explicitly  all singularities of the integrand in
the complex plane of the Mellin variable~$s$. Consequently,
integration in  (\ref{A26}) can be straightforwardly performed by
closing the integration contour in the left ($r<1$) or   right ($r>1$)
semiplane. The analytical expressions for $A_{2}^{(6),(l l) }(r<1)$ and
$A_{2}^{(6),(l l) }(r>1)$ turn to be two branches of one analytical function
$A_{2}^{(6),(l l) }(r)$
valid  in the whole interval $r \in (0,\infty)$,
\ba   && A_{2}^{(6),(l l) }(r)= \frac{2}{3}\; \left({\frac{1}{3}}-4\,
r^2 + 5\, r^4 - {\frac{16}{15}}\, r^6 \right)
 \left[-{\rm
Li_2} \left( \frac{1-r}{1+r} \right)+ {\rm Li_2} \left( -
\frac{1-r}{1+r} \right) +
 \right.\nonumber \\ &&
\left.
  \frac{1}{12}\;\pi^2 - 2\; {\rm Li_2}
\left(1-\frac{1}{r} \right)\right] \,
 +\frac{8}{3}\,\left[{\rm
Li_3}\,({r}^{2})- \bigg( {\rm Li_2}\,
({r}^{2}) +\frac{1}{3}\;\pi^2 \bigg)\ln (r) -\frac{2}{3} \;\ln^3 (r) \right] r^4  - \nonumber
\\ &&
\frac{8}{45}\left[{\rm Li_2}\left(\frac{1-r}{1+r} \right)-{\rm
Li_2}\left(- \frac{1-r}{1+r} \right)+\frac{1}{4}\;\pi^2\right] r + \frac{317}{324}+\frac{25}{27}\, \ln
(r)  -\frac{191}{45}\;r^2  -\frac{254}{45}\,r^2\ln \,(r) +
\nonumber \\ &&
\frac{16}{45}\, r^4 + \frac{32}{45} \, r^4 \ln
(r) .
 \label{A26LT1}
 \ea

Notice that Eq.~(\ref{A26LT1}) determines also  the mass-independent coefficient
$A_1^{(6)}$ in (\ref{A1}) as
 \ba
 A_1^{(6)} =\lim_{r\to 1} A_{2}^{(6),(l l) }(r)=
 -\frac{943}{324} - \frac{4 \pi^2}{135} + \frac{8}{3}\;\zeta(3)  \label{lim1}
 \ea
in agreement with the well-known result
(see Refs.~\cite{Jegerlehner:2017gek,Samuel-n-bubble} for
more details). Above, in Eq.~(\ref{lim1}),   $\zeta(3)$ denotes the  Euler-Riemann zeta function.

\subsection{Two different loops $(L\ell)$, Fig.~\ref{Fig-2-loop},  left panel }

For  diagrams of   type $( L\ell)$   one has $p=1$, $j=1$ and the corresponding
 Mellin--Barnes representation looks like
\ba &&
a_L(1,1)=-\frac{\alpha}{\pi}
\frac{2}{2\pi i}
\int\limits_{c-i\infty}^{c+i\infty}  ds
 \left(  4r^2 \right)^{-s}
\Gamma(s)\Gamma(1-s) \left(\frac{\alpha}{\pi}\right)  \Omega_1(s)\, \left(\frac{\alpha}{\pi}\right)  R_1(s), \label{fin3-differ} \ea
where the quantity $R_1(s)$ has already been computed, cf.
Eq.~(\ref{R1s}), whereas   $\Omega_1(s)$, Eq.~(\ref{Omp}), with the
polarization operator (\ref{lashk}), reads as
\begin{eqnarray} \label{om1}&&
\Omega_1(s)= \int_{0}^{1}   dx \;  x^{2s}(1-x)^{1-s}f(x) =
\bigg[\frac{(1-2 s) (1+2 s) \left(36+54
s-29 s^2-34 s^3+5 s^4+4 s^5\right)}{9(1-s) s (1+s) (2+s)}
 \nonumber \\ &&
 -2\pi
\left(1+s-s^2\right) \text{cot}(\pi  s)
\bigg]
\bigg(
\frac{2^{2s-1}}{\sqrt{\pi }}\bigg)
{\Gamma\bigg(-2-s\bigg)}{\Gamma\left(-\frac{1}{2}+s\right)}.
\label{Omega-1}
\ea
It is worth noting   that  for $\Omega_1(s=0)$ one has
\ba
-\Omega_1(0)=-\lim_{s\to 0} \Omega_1(s) =  \frac{119}{36}-\frac{\pi^2}{3} \, ,
\ea
which is exactly the limit (\ref{limA1}) for the coefficient  $A_1^{(4)}$ in (\ref{A1}).

By taking into account the singularities in Eqs.~(\ref{R1s})
and (\ref{Omega-1} ) in the complex plane of the variable~$s$,
 the Cauchy residue theorem provides the following explicit expression for $ A_{2}^{(6),( L\ell) }(r)$
 \begin{eqnarray}
  && A_{2}^{(6),( L\ell) }(r)=
 \left( \frac {4}{45\,{r}^{2}}  +\frac19+\frac43\,{r}^{2} - \frac {13}{9}\,r^4 \right)
    \bigg[ {\rm Li_2} \left( \frac{1}{{r}^{2}} \right) -2\ln \left(1- \frac{1}{{r}^{2}}\right)\ln(r) \bigg]- 2\left( 1+{r}^{4}\right) \times
   \nonumber
\\ &&
     {\rm Li_3} \left( \frac{1}{{r}^{2}} \right) +
\bigg( \frac1r+\frac23\,r+\frac{11}{3}\,r^3  \bigg)
\bigg\{{\rm Li}_3
\left(\frac{1}{r}\right)-{\rm Li}_3
\left(-\frac{1}{r} \right) +
 \bigg[{\rm Li}_2
\left(\frac{1}{r}\right)-{\rm Li}_2
\left(-\frac{1}{r} \right)
 \bigg] \ln(r)
 +
  \nonumber
\\ &&
 \frac12\big[ \ln(1+r)- \ln(1-r) \big]
 \ln^2(r)\bigg\}
    - \frac {8}{3}\;(1+r^4)
  \bigg[ {\rm Li_2} \left( \frac{1}{{r^2}} \right) - \frac12 \ln\left(1-\frac{1}{r^2} \right)\ln(r)\bigg]\ln(r) -
   \nonumber \\ &&
 \frac {8}{45}\;r^3 \bigg[
{\rm Li_2} \left( \frac{1}{r} \right)- {\rm Li_2} \left( -\frac{1}{r}
\right) + \big[ \ln \left( 1+r \right)-  \ln \left(1- r \right) \big]\ln(r) \bigg]-  \bigg(\frac {8}{45 r^2}+\frac{11}{9}+\frac{7}{3}r^2 \bigg)\times
 \nonumber
\\ &&
\ln^2(r)-\frac{
1853}{810} -\frac{349}{135} \ln(r) -\frac {53}{15} \, r^2 - \frac{64}{45}\,r^2 \ln(r) -{\frac {4\,{\pi}^{2}}{135\,{r}^{2}}} \, ,
  \label{lLL}
\end{eqnarray}

\noindent
which  defines  the single analytical function   $A_{2}^{(6),(l L)}(r)$   valid for all $r\in (0,\infty)$.
 As in the  case of
Eqs.~(\ref{4thorderLt1}) and~(\ref{4thorderGt1}), one should treat
with caution the complex logarithms in the square brackets  in (\ref{lLL}).

\section{Three lepton loops}\label{Sec:three}

In this section, we present  the details of calculations of the
8-th order corrections from the bubble-type diagrams, i.e.,
calculations of the coefficients $A_2^{(8)}(r)$ in (\ref{A23}). The
corresponding diagrams are depicted in Fig.~\ref{Fig-3-QFTHEP} for
which the ingredients $\Omega_p(s)$ and $R_j(s)$, Eqs.(\ref{Omp}) and
(\ref{Rj}) are: $p=1,\ j=2$ (left panel),  $p=2,\ j=1$ (central panel)
and  $p=0,\ j=3$ (right panel). The quantities $\Omega_0(s)$,
$\Omega_1(s)$, $R_1(s)$ and $R_2(s)$ have been already calculated in
Eqs.~(\ref{Omp0}), (\ref{Omega-1}), ~(\ref{R1s}) and (\ref{R2}),
respectively. The remaining $\Omega_2(s)$ and $R_3(s)$ will be
presented explicitly as they appear in  calculations of the
corresponding diagrams.
\begin{figure}[h]
\vspace*{2mm}
\includegraphics[width=0.24\textwidth]{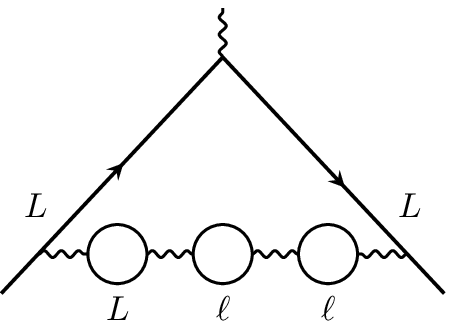}\hspace*{1cm}
\includegraphics[width=0.24\textwidth]{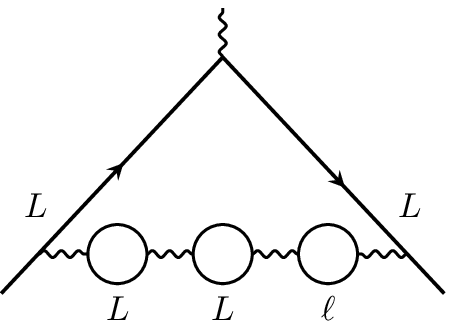}\hspace*{1cm}
\includegraphics[width=0.24\textwidth]{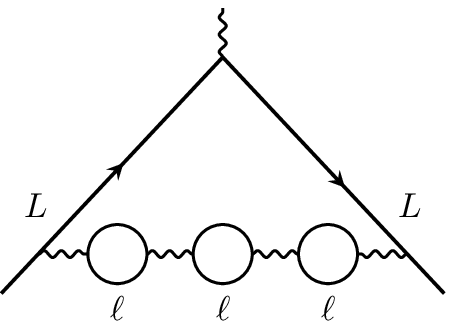}
%\vspace*{2mm}
\caption{The 8-th order diagrams considered in the present paper.
From left to right:
 one lepton loop as the external lepton  and two loops with $\ell\neq L$
($(L\ell\ell) $);
two lepton loops with the same leptons as
 the external one and one loop $\ell\neq L$  ($(LL\ell) $);
 three identical lepton lops $\ell\neq L $, ($(\ell\ell\ell) $). The relevant quantities
 $\Omega_p(s)$ and $R_j(s)$, Eqs.(\ref{Omp}) and (\ref{Rj}) are:
 $p=1,\ j=2$ (left panel),  $p=2,\ j=1$ (central panel) and  $p=0,\ j=3$ (right panel).
  }
\label{Fig-3-QFTHEP}
\end{figure}
\subsection{ Two loops with leptons  $(\ell \ell)\neq L$  and one loop with lepton  $L$:
$p=1$, $j=2$, left panel in Fig.~\ref{Fig-3-QFTHEP}}

The Mellin-Barnes integral representation for the contribution to the lepton anomaly from the diagram shown
in Fig.~\ref{Fig-3-QFTHEP},~left panel, reads as
\ba
 a_L(1,2)= \frac{\alpha}{\pi}
\frac{3}{2\pi i}
\int\limits_{c-i\infty}^{c+i\infty} \ ds
 %\left(\frac{\alpha}{\pi}\right)^{3}
 \left(  4r^2 \right)^{-s}
\Gamma(s)\Gamma(1-s) \left(\frac{\alpha}{\pi}\right) \Omega_1(s)\ \left(\frac{\alpha}{\pi}\right)^2 R_2(s) ,
\label{aLtwo}
\ea
where  $R_2(s)$ and $\Omega_1(s)$  are given by Eqs.~(\ref{R2}) and (\ref{Omega-1}), respectively.
From these expressions one can infer explicitly the locations of the singularities of the integrand
(\ref{aLtwo}) contributing to  the Cauchy residue theorem in the complex plane of the Mellin variable~$s$. Closing the
integration contour in the left semiplane $(r<1)$ and computing the corresponding residues,    we obtain
\ba &&
 A_{2}^{(8),(L\ell\ell ) }(r<1)=\nonumber
 \frac{16}{5r}\left( -{\frac {3}{7}}+ \frac {1 }{9} \; r^2 \right)
 \bigg[
\big[  {\rm Li}_2\left(r \right) -{\rm Li}_2\left(-r \right) \big] \ln ( r  ) -
\frac12 \big[  \,\ln  \left(1+r\right) - \,\ln  \left( 1-r \right)
\big]
 % 2
\nonumber \\ &&
\times  \ln^2 (r)  + {\rm Li}_3\left(-r \right) -{\rm Li}_3\left(r \right)  \bigg] -
 \left(  \frac{13}{9} -4\,{r}^{2}+
\frac {67\,}{27}\,{r}^{4} + \frac {2 }{27}\,{r}^{6}  \right) \bigg[{
\rm Li}_2\left({r}^{2} \right) +2\,\ln ( 1 -r^2
)\ln  \left( r \right)
 % 3
 \nonumber \\ &&
 - \frac13\,{\pi}^{2}\bigg] +
 \left(\frac13+4\,{r}^{2}- \frac {73\,}{15}\,{r}^{4} + \frac {88}{315}\,{r}^{6}  \right) {\rm Li}_3 ({r}^{2}) -
\frac23  \bigg[ \left( \frac23+8{r}^{2}- \frac {154}{15}\,{r}^{4} + \frac
{88}{105}\,{r}^{6}  \right) {\rm Li}_2 ({r}^{2} )
 \nonumber
 \ea
 \vspace*{-9mm}
 \ba &&
+ \left(
\frac19+\frac43{r}^{2}-{\frac {13}{9}}\,{r}^{4}
 \right)\pi^2 \bigg]  \ln ( r )
  + 4\left( 1+2r^4
\right) \bigg[
 {\rm Li}_4\left({r}^{2} \right)-  {\rm Li}_3\left({r}^{2} \right)  \ln ( r )
 +\frac13 \bigg( {\rm Li}_2\left({r}^{2} \right)  -
\nonumber \\ [0.1cm] &&
\frac{1}{3} \pi^2 \bigg)  \ln ^2 (r) \bigg] +  \left( {\frac {9911}{2835}}+{\frac {9136}{945}{r}^{2}}-{\frac {668}{945}\,{r}^{4}} \right) \ln  \left( r \right) +
\bigg[ \frac{2869}{945} - \frac {58}{35}\,{r}^{2} +
\frac {4162}{945}\,{r}^{4} + \frac {4 }{27}\,{r}^{6} -
\nonumber \\ [0.1cm]&&
\frac23\left(\frac13 +4 \,{r}^{2}- \frac {89 }{15} {r}^{4}+
 \frac {88\,}{105}  {r}^{6} \right) \ln  (1-r^2 ) \bigg]  \ln^2  ( r )-
\frac{32}{45}\left( 1 - \frac {11}{41}\,{r}^{2}  \right) {r}^{4}  \ln^3  ( r ) +
\nonumber  \\ [0.1cm] &&
{\frac{80321}{68040}}+ \frac {6509}{630} \,{r}^{2}
  - \frac {334 }{945}\, {r}^{4} -\frac{4}{45} \bigg( 1+2r^4\bigg) {\pi}^{4} ,
 \label{llL}
\ea
where, regardless that the integral (\ref{aLtwo}) has been calculated by the Cachy's
theorem in the left semiplane,   the obtained expression represents the sought
analytical function $A_{2}^{(8),(L\ell\ell ) }(r)$ valid in the whole interval  $r\in (0,\infty)$. This assertion has
been checked by calculating the integral closing the contour in the
right semiplane and also by direct numerical calculations of the
integral (\ref{double}) with the polarization operator~(\ref{lashk}).

\subsection{ Two leptons $(L L)$ as the external one and one loop with
$l\neq L$: $p=2$, $j=1$, central panel in Fig.~\ref{Fig-3-QFTHEP} }

For these diagrams the Mellin-Barnes representation provides
\ba
 a_L^{  LL\ell}(2,1)= \frac{\alpha}{\pi}
\frac{3}{2\pi i}
\int\limits_{c-i\infty}^{c+i\infty} \ ds
 \left(  4r^2 \right)^{-s} \Gamma(s)\Gamma(1-s)\left(\frac{\alpha}{\pi}\right)^{2} \Omega_2(s)\,
 \left(\frac{\alpha}{\pi}\right)  R_1(s) ,
\label{aL3} \ea
where $R_1(s)$ is given by Eq.~(\ref{R1s}). Due to
presence of the squared  polarization operator and logarithmic
functions in  Eqs.~(\ref{Omp}) and (\ref{reP1}), calculations of
$\Omega_2(s)$ turn to be rather cumbersome.   To present the results in a
more or less compact  form, we introduce several auxiliary functions
containing integrals with powers of the logarithm $\ln^k (1-x)$, $k=0,~1,~2$:
\begin{equation}
X_k(s,n)=\int\limits_0^1  dx \; x^{2s+n} (1-x)^{1-s}\ln^k(1-x).
\end{equation}
Then we obtain

\vskip -0.2cm \noindent
\ba
&& X_0(s,n) =
\frac{\Gamma(2 - s) \Gamma(1 + n + 2 s)}{\Gamma(3 + n + s)} \, ,   \quad X_1(s,n)=X_0(s,n)
\bigg( \psi \left( 2-s \right) -\psi \left( 3+n+s \right) \bigg)  , \quad \nonumber
 \\ [-0.2cm]
 &&
 \label{X012}
 \\
 &&
X_2(s,n)=
 X_0(s,n)
\left[   \bigg( \psi \left( 2-s \right)  -   \psi \left( 3+n+s \right)  \bigg) ^{2}+ \;
\psi^{(1)} \left(2-s \right) - \; \psi^{(1)} \left(3+n+s\right)  \right],
\quad
 \nonumber
\ea

\noindent
where $\psi(x)$ and $\psi^{(1)}(x)$ are the  polygamma functions of the   order $0$ and $1$, respectively.
Then, with (\ref{X012}),  equation (\ref{Omp}) casts the form
\ba
&& \Omega_2(s)\;  = \;
\frac{25}{81} X_0(s, 0) + \frac{16}{9}  X_0(s, -4) - \frac{32}{9} X_0(s, -3) +
 \frac{8}{27}  X_0(s, -2) + \frac{40}{27}  X_0(s, -1) -
  \nonumber \\ [0.1cm] &&
  \frac{10}{27} X_1(s, 0) +
 \frac{32}{9}  X_1(s, -5) - \frac{80}{9}  X_1(s, -4) + \frac{104}{27}  X_1(s, -3) +
\frac{ 28}{9}  X_1(s, -2) -  \frac{8}{9}  X_1(s, -1) +
 \nonumber \\ [0.1cm] &&
 \frac{1}{9} X_2(s, 0) +\frac{16}{9}  X_2(s, -6) -
 \frac{16}{3}  X_2(s, -5) + 4 X_2(s, -4) +
 \frac{8}{9}  X_2(s, -3) - \frac{4}{3}  X_2(s, -2).
\label{om2}
\ea
It immediately  follows from Eq.~(\ref{om2})  that
\ba
A_1^{(6)}= \Omega_2(0) = -\frac{943}{324}-\frac{4\pi^2}{135}  + \frac{8}{3}\,\zeta(3).
\label{A16}
\ea
Having calculated explicitly $\Omega_2(s)$ and $R_1(s)$,   further integration in (\ref{aL3}) is
 performed
by the Cauchy residue theorem. Closing the integration contour in the left semiplane ($r<1$) or right
($r>1$) semiplanes
in (\ref{aL3}), we obtain the coefficient corresponding to the anomaly $ a_L^{LL\ell}~$.

\subsubsection{ Closing the integration contour in the left semiplane of $s$: $A_2^{(8),(LL\ell)}(r<1)$}

The result of integration for $A_2^{(8),(LLl)}(r<1)$  can be  represented  in the following form:
 \ba &&
A_2^{(8),(LL\ell)}(r<1)=
\frac43 \;\bigg\{
{\frac {34}{105}}+{\frac {4}{35\,{r}^{2}}}+{\frac {323}{
210}}\,{r}^{2}+{\frac {89}{36}}\,{r}^{4}
+\frac{2}{105}(21r-r^3) \big[ \ln \left( 1+r \right)-  \ln \left(1- r
\right)  \big]
%2
\nonumber \\ &&
+ \left( \frac16+{\frac {4}{35\,{r}^{4}}}+{\frac {4}{15\,{r}^{2}}}+2\,
{r}^{2}-{\frac {13}{6}}\,{r}^{4} \right)
 \ln  \left( 1-{r}^{2} \right)-\left( 2+{r}^{4} \right) {\rm Li}_2 (r^2)
 \bigg\}\ln^2(r) + 2 \;\bigg\{{\frac {1813}{43200\,{r}^{4}}} -
 %3
  \nonumber \\ &&
{\frac {8731477}{4536000}} -{\frac {371429}{907200\,{r}^{2}}}+{\frac {33907}{60480}}\,{r}^{2}+{\frac {590}{189}}\,{r}^{4} +
   \left( \frac15 +{\frac {8}{105\,{r}^{4}}}+{\frac {8}{45\,{r}^{2}}}+\frac83\,{r
}^{2}-{\frac {26\,{r}^{4}}{9}} \right) {\rm Li}_2 (r^2)+
 %3a
  \nonumber \\ &&
\left( \frac83+2\,{r}^{4} \right)  {\rm Li}_3 (r^2)+
\left( -{\frac {37939}{69120}}+{\frac {7}{1536\,{r}^{4}}}-{\frac {173}{4608\,{r}^{2}}}+
{\frac {55}{1536}}\,{r}^{2}-{\frac {442
}{567}}\,{r}^{4} \right) {\pi }^{2}  - \frac43\, \left(2+{r}^{4} \right) \times
%4
 \nonumber \\ [0.2cm] &&
\zeta  \left( 3 \right)  \bigg\} \ln(r) -
%5
   \bigg[\; {\frac {1813}{43200{r}^{5}}}\, -{\frac {25}{72\,{r}^{3
}}}-{\frac {145}{32\,r}}+{\frac {1481}{192}}\,{r}^{3}+
\left( {\frac {7}{1536\,{r}^{5}}}-{\frac {5}{128\,{r}^{3}}
}-{\frac {145}{256\,r}}-{\frac {167}{384}}\,r- \right.
 %6
  \nonumber \\ [0.2cm]
  &&
  \left.
{\frac {1481}{
1536}}\,{r}^{3} \right){\pi }^{2}
\bigg]  \bigg[ {\rm Li}_2 (-r)-{\rm Li}_2 (r) +
\big[ \ln \left( 1+r \right)-  \ln \left( 1-r
\right) \big]\ln(r)
 \bigg] +{\frac {16}{15}}\left(r-{\frac {{r}^{3}}{21}} \right)  \bigg[ {\rm Li}_3 (r) -
 %7
  \nonumber \\ [0.1cm] &&
{\rm Li}_3 (-r)-\big( {\rm Li}_2 (r)-{\rm Li}_2 (-r) \big) \ln(r)
 \bigg] -
\left( {\frac {13}{45}}+{\frac {8}{105\,{r}^{4}}}+{\frac {8}{45\,{r}
^{2}}}+4\,{r}^{2}-\frac{13}{3}\,{r}^{4} \right)
{\rm Li}_3 \left({r}^{2} \right) +
%8
  \nonumber  \\ [0.2cm]&&
 \left( {\frac {5383}{1350}}-{\frac {28}{9}}\,{r}^{2}- {\frac {125}{54}}\,{r}^{4}+\frac23 \,{\pi }^{2}+\frac13 \,{\pi }^{2}{r}^{4} \right) \bigg[ {\rm Li}_2 \left( {r}^{2} \right) +2 \ln(1-r^2)\ln(r) \bigg]- 4\left( 1+{r}^{4} \right) {\rm Li}_4  \left({r}^{2}\right)
    \nonumber \\ [0.2cm]
    &&
    %9
 - {\frac {1813}{21600\,{
r}^{4}}}+{\frac {1035989}{~1360800\,{r}^{2}}} +{\frac{519835531}{34020000}}+ \,{\frac {147587}{30240}}\,{r
}^{2}- {\frac {~558857}{59535}}\,{r}^{4}  +
\left( {\frac {211546\,{r}^{4}}{178605}}+{\frac {6799\,{r}^
{2}}{80640}} \right.
 \nonumber \\ [0.2cm] &&
 \left.
 -{\frac {97463}{518400}}+{\frac {533}{6912}}\,{r}^{2} -{
\frac {7}{768}}\,{r}^{4} \right) {\pi}^{2} +
\frac{\pi^{4}}{27}(2+{r}^{4}) -  \left({\frac {2}{45}}+\frac83\,{r}^{2}-{\frac {676}{189}} \,{r}^{4}
 \right) \zeta(3)
+ S_1(r)
  \label{Adva}
%\label{LLl-A}
\ea

\noindent
with
%\vskip -0.2cm \noindent
\begin{eqnarray}  \label{sumlt1}
&& S_1(r) = 2 \sum_{n=3}^{\infty}
\bigg[ C_1(n)\; \bigg( \psi^{(2)}({n})+ \psi^{(1)}({n})\ln( r^2) \bigg) - C_2(n)\;\psi^{(1)}({n}) \bigg]  r^{2n}\,,  \\[0.2cm]
&& C_1(n)=(-60+5 n+86 n^2+36 n^3+3 n^4) /Y_1(n) \,, \nonumber \\
 && C_2(n)=
\left(5400-1080n-67455n^2+20260n^3+107834 n^4-17494n^5\right. \nonumber    \\
&&\left. ~~~~~~~~ -77654n^6-10816n^7+14768n^8 +4704 n^9+288 n^{10}
\right)/[Y_1(n)]^2  \, , \nonumber  \\
&&Y_1(n)=  n
(n-2)(2n+5)(4n^2 - 9) (4n^2-1)  \,. \nonumber
\end{eqnarray}

A scrupulous analysis of the sum $S_1(r<1)$ demonstrated that
it converges very fast in the region $r<1$ and can be calculated  numerically  with any desired accuracy. However,  we  have not succeeded in finding explicitly its convergency function and hence presenting
 $ A_2^{(8),(LL\ell)}(r<1)$ in terms of  a single complex function, similar to Eq.~(\ref{llL}).
Moreover, as it can be easily seen,  the sum  $S_1(r)$ diverges for $r>1$.  Consequently,
 $ A_2^{(8),(LL\ell)}(r<1)$, Eq.~(\ref{Adva}), cannot be straightforwardly  continued analytically to the right semiplane, i.e., the resulting expressions must be considered
   strictly as only the left semiplane branch of the analytical function $A_2^{(8),(LL\ell)}(r)$.

\subsubsection{Closing the integration contour in the right semiplane of $s$: $A_2^{(8),(LL\ell)}(r>1)$}

\vskip -0.2cm \noindent
By doing as before, we get
\ba &&\label{Advaright}
A_2^{(8),(LL\ell)}(r>1) =\frac{16}{9} \big( 2+r^4\big)\,  \ln  ^{4} ( r  ) +
  \bigg(
 \frac {1949}{1728} -  \frac {701\,}{192}{r}^{2}  +{\frac {52\,}{9}}{r}^{4}
 - \frac {1447}{6720\,{r}^{4}} -{\frac {359}{960\,{r}^{2}}}
  \bigg)   \ln^3  ( r ) +
% 2
\nonumber \\ &&
  \bigg[ \left( \frac29+\frac83\,{r}^{2}
  -  \frac {26 }{9} r^{4} + \frac {16}{105\,{r}^4}+ \frac {16}{45\,{r}^{2}} \right) \ln  \left( {r}^{2}-1 \right) + \frac{4}{9} \big(2+r^4\big) \pi^2+
  \frac{11261}{2240}+ \frac {295409}{20160}\,{r}^{2} +
\nonumber \\ &&
% 3
 {\frac {268}{27}}\,{r}^{4}+ {\frac {7}{192\,{r}^{4}}}-{\frac {9439}{60480\,{r}^{2}}}
  \bigg]  \ln^2( r)+  \bigg\{  \bigg[
 - \frac{2}{9} \big(2+r^4\big) \pi^2- {\frac {56\,{r}^{
2}}{9}}-{\frac {134\,{r}^{4}}{27}}-{\frac {16}{27}}  \bigg] \ln
 \left( {r}^{2}-1 \right) -
   \nonumber \\ &&
%4
\left( \frac49+ \frac{16}{105\,{r}^{4}}+{\frac {16}{45\,{r}^{2}}}+\frac{16}{3}\,{r}^{2}-{\frac {52
\,{r}^{4}}{9}} \right) {\rm Li_2} \left(\frac{1}{{r}^{2}} \right) +
\left( {\frac {2717}{6912}}+{\frac {173}{6912\,{r}^{2}}}+{\frac {323
}{768}}\,{r}^{2}-{\frac {7}{2304{r}^{4}}\,}  \right) {\pi }^{2}
  \nonumber \\ [0.2cm]&&
% 5
+{\frac {7}{864{r}^{4}}\,} -{\frac {1291}{10080\,{r}^{2}}}+{\frac {501857}{30240}}\,{r}^{2} +{
\frac {380971}{30240}}
  \bigg\}\ln(r) + \bigg[
\frac{\pi^2}{9}\; \big(2+r^4\big)   +
 {\frac {8}{27}}+{\frac {28}{9}}\,{r}^{2}+{\frac {67}{27}}\,{r}^{4}  \bigg]
 %6
    \nonumber \\ [0.2cm] &&
\times \,{\rm Li_2} \left(\frac{1}{{r}^{2}} \right) - \bigg(
 \frac13+{\frac {8}{105\,{r}^{4}}}+{\frac {8}{45\,{r}^{2}}}+4\,{r}^{2}-\frac{13}{3}
\,{r}^{4} \bigg)
{\rm Li_3} \left(\frac{1}{{r}^{2}} \right)+ \frac{14}{3} \; \big( 2+{r}^{4} \big)
{\rm Li_4} \left(\frac{1}{{r}^{2}} \right) +
%7
\nonumber \\ &&
\bigg[ {\frac {7405\,{r}^{3}}{864}}+{\frac {167}{48}}\,r-{\frac {5}{16\,{r}^{3}}}+{\frac {35}{864\,{r}^{5}
}} + \bigg(
{\frac {5}{384\,{r}^{3}}}+{\frac {145}{768\,r}}-{\frac {7}{4608\,{r}^{
5}}}+{\frac {1481}{4608}}\,{r}^{3}+{\frac {167}{1152}}\,r
 \bigg)\pi^2 \bigg] \times
   %8
     \nonumber \\[0.2cm] &&
     \bigg[ {\rm Li_2} \left(-\frac{1}{{r}} \right) -
{\rm Li_2} \left(\frac{1}{{r}} \right) + \bigg( \ln  \left(1-\frac{1}{{r}} \right)-\ln  \left(1+\frac{1}{{r}} \right)
 \bigg)\ln(r)  \bigg]
 +\bigg( {\frac {7}{1152\,{r}^{5}}}-{\frac {5}{96\,{r}^{3}}}-{\frac {145}{192\,
r}}-
   %9
     \nonumber \\ [0.2cm]&&
{\frac {167}{288}}\,r- {\frac {1481}{1152}}\,{r}^{3}  \bigg)
\bigg\{\bigg[ \ln  \left(1+\frac{1}{{r}} \right)-\ln  \left(1-\frac{1}{{r}} \right)
 \bigg]\ln^{3}(r)-3\; \bigg[{\rm Li_2} \left(-\frac{1}{{r}} \right) - {\rm Li_2} \left(\frac{1}{{r}} \right)
 \bigg] \ln^{2}(r)   -  \nonumber
 \ea
 \vspace*{-8mm}
 \ba
   %10
  &&
 6\; \bigg[ {\rm Li_3} \left(-\frac{1}{{r}} \right) - {\rm Li_3} \left(\frac{1}{{r}} \right) \bigg] \ln(r)   - 6 \; {\rm Li_4} \left(-\frac{1}{{r}} \right) + 6\;{\rm Li_4} \left(\frac{1}{{r}} \right)  \bigg\}+ \bigg({\frac {8}{15}}\,r-{\frac {8}{315}}\,{r}^{3} \bigg) \times
  %11
     \\ [0.2cm] &&
 \bigg\{ \bigg[ \ln  \left(1+\frac{1}{{r}} \right)-\ln  \left(1-\frac{1}{{r}} \right)
 \bigg]\ln^{2}(r)-2\; \bigg[{\rm Li_2} \left(-\frac{1}{{r}} \right) - {\rm Li_2} \left(\frac{1}{{r}} \right)
 \bigg] \ln(r)   - 2\;
{\rm Li_3} \left(-\frac{1}{{r}} \right) +
  %12
     \nonumber \\ [0.2cm] &&
     2\;  {\rm Li_3} \left(\frac{1}{{r}} \right) \bigg\}  + \frac83 \, \big( 2+r^4\big)\,
  \bigg\{2\; \bigg[{\rm Li_3} \left(\frac{1}{{r^2}} \right) + {\rm Li_2} \left(\frac{1}{{r^2}} \right)\bigg]\ln(r) -\frac13 \ln(r^2-1)\ln^2(r)
 \bigg\}\ln(r)
  +
   \nonumber \\ [0.2cm] &&
   %13
\bigg( {\frac {6917}{20736}}-{\frac {533}{20736\,{r}^{2}}}+{\frac {7}{2304\,{r}^{4}}} +{\frac {1225}{2304}}\,{r}^
{2} \bigg)\pi^2
-{\frac {7}{864\,{r}^{4}}}+{\frac {2059}{30240\,{r}^{2}}}+{\frac {
23731\,{r}^{2}}{1120}}+{\frac {1208147}{90720}}+
   \nonumber %\\ [0.4cm] &&
   %14
S_2(r) ,
\ea
where
\begin{equation}\label{sumgt1}
S_2(r) = 2 \sum_{n=1}^{\infty}
\bigg[ C_1(-n)\; \bigg( \psi^{(2)}({n})+ \psi^{(1)}({n})\ln( r^2) \bigg) - C_2(-n)\;\psi^{(1)}({n}) \bigg]  r^{-2n}\,.
\end{equation}

As in  the case $r<1$, the sum (\ref{sumgt1})  converges fast  in
the interval $r\in [1,\infty)$ and diverges  at $r<1$. Also,
we have not found explicitly the convergency function for  $S_2(r)$,
so that a direct analytical continuation   to the left semiplane
remains hindered. Correspondingly,
Eqs.~(\ref{Advaright})-(\ref{sumgt1}) define only the right semiplane
branch of the analytical function $A_2^{(8),(LLl)}(r)$.

\subsection{ Three identical leptons $(lll)$, $l\neq L$: right panel of Fig.~\ref{Fig-3-QFTHEP}, $p=0$, $j=3$}

For these diagrams the lepton anomaly reads as
\ba
 a_L^{\ell\ell\ell}(0,3)= \frac{\alpha}{\pi}
\frac{1}{2\pi i}
\int\limits_{c-i\infty}^{c+i\infty} \ ds
 \left(  4r^2 \right)^{-s}
\Gamma(s)\Gamma(1-s) \Omega_0(s)\,  \left(\frac{\alpha}{\pi}\right)^{3} R_3(s) , \label{Fig3},
\ea
where $\Omega_0(s)$ is defined in Eq.~(\ref{Omp0}). As for $R_3(s)$,
direct calculations by  Eqs.~(\ref{Rj1}) and (\ref{reP1})  provide
\begin{equation}
R_3 (s) = \frac{\sqrt{\pi}}{864}\frac{\Gamma(s)}{\Gamma
\left(\frac{11}{2}+s\right)}
\left\{\frac{P(s)}{s(1+s)(2+s)}-27 (1+s)(35+21s+3s^2)\left[ \pi^2-6\
\psi^{(2)}(s)\right] \right\}, \label{MMB-3-loop-R3}
\end{equation}
%\vskip 0.2cm \noindent
where $P(s)=3492 - 8748 s - 26575 s^2 - 9214 s^3 + 18395 s^4 + 17018
s^5 + 5120s^6 + 512s^7\,$.

Explicitly, the coefficient $A_{2}^{(8),\ell\ell\ell}(r)$ is
\begin{eqnarray}\label{MB-3-loop-3Afull}
& & A_{2}^{(8),(\ell\ell\ell)}(r) = \frac{1}{2\pi
i}\int\limits_{c-i\infty}^{c+i\infty }  ds  \; r^{-2s} \left[
\frac{(1-s)}{(s+2) (2s+1)(2s+3)(2s+5)(2s+7)(2s+9)} \right] \times \nonumber \\
[0.2cm]   & &
 \bigg[ \frac{-(1-s)
(-3492+5256s+31831s^2+41045s^3+22650s^4+5632s^5+512s^6)}{27s(s+1)^2(s+2)}
  ~~~ \nonumber\\
 & &  -\; {\pi^2(35+21s+3s^2)} +\;
{6\psi^{(1)}(s)(35+21s+3s^2)}\bigg ] \frac{\pi^2}{\sin^2(\pi s)} \; .
\end{eqnarray}
Equations~(\ref{Omp0}), (\ref{MMB-3-loop-R3}) and (\ref{MB-3-loop-3Afull}) define explicitly all the singularities in the complex plane of the variable $s$. Accordingly, the integration contour can be closed in the left ($r<1$) or right ($r>1$) semiplane of~$s$:

\vskip -0.2cm \noindent

\subsubsection{Integration contour in the left semiplane: $A_2^{(8),(\ell\ell\ell)}(r<1)$ }

\vskip -0.2cm \noindent
\ba &&
A_2^{(8),(\ell\ell\ell)}(r<1) = \bigg[
 -{\frac {8609}{5832}}+ \frac {1760 }{567}{r}^{2} + \frac {7899407}{396900}\,{r}^{4} -
  \frac {692549}{1786050}\,{r}^{6} - \frac {50171}{793800}\,{r}^{8} +\nonumber \\ &&
\frac13\, \left( - \frac {25}{54} + \frac {4237}{1152}\,{r}^{2} - \frac {13709}{3456}\,{r}^{4} +
 \frac {767}{1152}\,{r}^{6} + \frac {11}{128} \,{r}^{8} \right) {\pi }^{2}+\frac15\,{\pi }^{4}{r}^{4}-
  \frac {1}{1536}{\pi }^{4}r{\it V_1(r)}+\nonumber \\ &&
\frac{r}{4}\, \left( \frac {101}{16} - \frac {1025}{18}\,{r}^{2} - \frac {1813 }{120}\,{r}^{4} -
 \frac {3229}{525}\,{r}^{6} - \frac {1292159}{1587600}\,{r}^{8} +{\frac {{\pi }^{2}}{192}}{\it V_1(r)} \right) \bigg( {\rm Li_2} \left( r \right) -{
\rm Li_2} \left(-r \right)  \bigg) +\nonumber \\ &&
 \frac {r}{64} {\it V_1(r)}\bigg(
{\rm Li_4} \left( r \right) -{\rm  Li_4} \left(-r \right)\bigg)
+{\it V_2(r)}\,{\rm Li_2} \left( {r}^{2} \right) -  \frac {32}{315} \,{r}^{2} \left( 7\,{r}^{2}-12 \right) {\rm Li_3} \left(  {r}^{2} \right)  -2\,r^4{\rm Li_4} \left(r^2\right) + \nonumber \\ &&
% !!!
 \left( -\frac29+ \frac {136}{35}\,{r}^{2} - \frac {64}{15}\,{r}^{4} +
  \frac {16}{15}\,{r}^{6}  \right) \zeta  \left( 3 \right)\bigg ] + 2\,\bigg[ - \frac {317}{324} + \frac {15863}{2835}\,{r}^{2} -
  \frac {12963943}{1587600}\,{r}^{4} + \frac {1046261}{2381400}\,{r}^{6} -
\nonumber \\ &&
  \frac {50171}{1587600}\,{r}^{8} +  \frac13\,{\pi }^{2} \left( -\frac19+ \frac {721}{768}\,{r}^{2} + \frac {2981}{2304}\,{r}^{4} +
   \frac {263}{768}\,{r}^{6} + \frac {11}{256}\,{r}^{8}  \right)
   +\frac12\,r \left( {\it V_3(r)}-\frac {{\pi }^{2}}{768}{\it V_1(r)} \right) \times
 \nonumber \\ &&
     \bigg( \ln  \left( r+1
 \right) -\ln  \left( 1-r \right)  \bigg) -  \frac {1}{128} {\it V_1(r)}\,r\bigg( {\rm Li_3} \left(r \right) -{\rm Li_3} \left( -r \right)  \bigg)
    +{\it V_2(r)}\,\ln ( 1-{r}^{2}) +\frac{16}{15}
   \times
 \nonumber \\ &&
    \bigg( -\frac {4}{7}+\frac{1}{3}\, r^2 \bigg) r^2{\rm Li_2} ({r}^{2} )
+2\,{r}^{4}{\rm Li_3}\left({r}^{2} \right) +
4\,\zeta  \left( 3 \right) {r}^{4} \bigg]\ln  ( r )+  4\, \bigg[ - \frac {25}{108} + \frac {4237}{2304}\,{r}^{2} - \frac {13709}{6912}\,{r}^{4} +
\nonumber \\ &&
\frac13\,{\pi }^{2}{r}^{4}+
\frac {767}{2304}\,{r}^{6} + \frac {11}{256}\,{r}^{8} +
 \frac { r  }{512}{\it V_1(r)}\bigg( {\rm Li_2} ( r  ) -{\rm Li_2} ( -r )  \bigg)-
 {r}^{4}{\rm Li_2} \left( {r}^{2}\right)  \bigg]    \ln ^{2} ( r  ) +\nonumber \\ &&
\bigg[ - \frac {4}{27} + \frac {721}{576}\,{r}^{2} + \frac {2981}{1728}\,{r}^{4} +
 \frac {263}{576}\,{r}^{6} + \frac {11}{192}\,{r}^{8}
 - \frac {r}{384}{\it V_1(r)}
 \bigg( \,\ln( r+1) - \,\ln( 1-r)  \bigg)   -\, \nonumber \\ &&
 \frac83\,{r}^{4}\ln (1 -{r}^{2})  \bigg]  \ln^{3} ( r )   +\frac43\,{r}^{4}  \ln^{4}( r ) +S_3(r)
 \label{lllLeft}
 \ea
with
\ba S_3(r) =6 \sum\limits_{n=4}^\infty
  \bigg[C_3(n)H^{(2)}(n)+C_4(n)\psi^{(2)}({n+1})-C_4(n)H^{(2)}(n)\ln(r^2)\bigg] r^{2n}  ,
 \label{s3lt1}
\ea
where $H^{(2)}(n)$ is the $n$-th  generalized harmonic number  and the shorthand notation is
\ba &&
 C_3(n)=
( -295995+836500\, n-787336\,n^2+206366\,n^3+131386\,n^4-
 114304\, n^5 +\nonumber \\ &&  \phantom{P_4(n)=(} 35216\, n^6-5088\, n^7+288\, n^8)/[Y_2(n)]^2 , \nonumber\\ &&
C_4(n)=(1+n) (35-21\, n+3\, n^2)/Y_2(n),\nonumber
, \\ &&
Y_2(n)=(n-2) (2\, n-9)(2\, n-7) (2\, n-5) (2\, n-3)(2\,n-1) ,\nonumber
\ea
and
 \ba  && V_1(r)= -101 +820 \, r^2 +210\,  r^4 + 84\, r^6 +11\, r^8 \\[0.2cm]  &&
 V_2(r)= -\frac{97}{2835} +\frac{10348}{11025}\, r^2+\frac{5524}{675} \,r^4-\frac23 \,\pi^2\, r^4 \\[0.2cm]  &&
 V_3(r)= -\frac{101}{64}  +\frac{1025}{72}\, r^2+\frac{1813}{480}\, r^4+
\frac{3229}{2100}\, r^6+\frac{1292159}{6350400} \,r^8,
 \ea

 \subsubsection{Integration contour in the right semiplane:
 $A_2^{(8),(\ell\ell\ell)}(r>1)$ }

 \vskip -0.2cm \noindent
\begin{eqnarray} \label{Fig3B-D0}
&& A_2^{(8),(\ell\ell\ell)}(r>1)
= \left[ \frac{307987}{893025}- \frac{1620853897}{222264000} r^2 -
\frac{137496269}{31752000} r^4 - \frac{30585647}{19051200} r^6 -
\frac{1292159}{6350400}r^8 + \right. \nonumber \\ [0.2cm]
 &&
\left. \left(-\frac{97}{2835}+ \frac{10348}{11025} r^2 +
\frac{5524}{675} r^4 \right)\ln\left(1- \frac{1}{r^2}\right)+\left(
\frac{64}{105}-  \frac{16}{45}r^2\right)r^2\;{\rm Li}_2
\left(\frac{1}{r^2}\right)
 \right] \ln(r^2)
 +\nonumber \\ [0.2cm]
 &&
\left(-\frac{101}{64}r+\frac{1025}{72}r^3+
 \frac{1813}{480}r^5 + \frac{3229}{2100}r^7
 +\frac{1292159}{6350400}r^9 \right )
\bigg[{\rm Li}_2
\left(\frac{1}{r}\right)-{\rm Li}_2
\left(-\frac{1}{r} \right)+ \ln (r)\times
\nonumber \\ [0.2cm]
 &&
\bigg( \ln \left( r+1 \right)-  \ln \left( r
-1 \right) \bigg)  \bigg]
+\left(\frac{97}{2835}-\frac{10348}{11025}r^2-\frac{5524}{675}r^4\right){\rm Li}_2 \left(\frac{1}{r^2}\right)+ \left(\frac{128}{105}-\frac{32}{45}r^2\right) \times
\nonumber \\ [0.2cm]
 &&
~~ r^2
{\rm
Li}_3\left(\frac{1}{r^2}\right)
+
\frac{415506937}{281302875}  -\frac{79924199429}{3889620000}r^2-\frac{628065509}{79380000}r^4-\frac{89172623}{28576800}r^6
 -\frac{1292159}{3175200}r^8 +
 \nonumber \\
 &&
~~ S_4(r)\, ,
\end{eqnarray}
where
\ba
S_4(r)=6 \sum\limits_{n=2}^\infty
  \bigg[C_3(-n)H^{(2)}_{n-1}-C_4(-n)\psi^{(2)}({n})-C_4(-n)H^{(2)}_{n-1}\ln(r^2)\bigg] r^{-2n} .
 \label{s4lg1}
\ea

Note that the sums $S_1(r)$, Eq.~(\ref{sumlt1}), and $S_3(r)$,
Eq.~(\ref{s3lt1}), converge rapidly at $r<1$ and diverge at $r>1$.
Contrarily, the sums $S_2(r)$, Eq.~(\ref{sumgt1}), and $S_4(r)$,
Eq.~(\ref{s4lg1}), converge   at $r>1$ and diverge at $r<1$. It means
that these sums can be applied  strictly in the region of their
convergence. However, an important comment is in line here. Namely,
we observed that despite the fact that the coefficient
$A_2^{(8),(\ell\ell\ell)}(r>1)$ has been obtained exclusively  for
$r>1$, its formal employment in the region  $r<1$,  from $r\ll 1$  up
to $r=0.5\div 0.55 $, shows that (\ref{Fig3B-D0}) provides
absolutely the same numerical results as the coefficient
$A_2^{(8),(\ell\ell\ell)}(r<1)$, Eq.~(\ref{lllLeft}). It means that
if one finds explicitly the convergence function for $S_4(r)$, it
would be possible to analytically continue
$A_2^{(8),(\ell\ell\ell)}(r)$ by a single function valid in the whole
interval $r\in (0,\infty)$. In the case when the results of integration do
not contain infinite sums, the analytical functions for one, two and
three loops are determined by Eqs.~(\ref{4thorderGt1}),(\ref{A26LT1}),
(\ref{lLL}) and (\ref{llL}), respectively.

\section{Discussions}\label{results}

With the above results we conclude our analytical consideration of the
radiative corrections to the lepton anomaly from QED diagrams with
insertions of the photon polarization operators up to the eighth
order. The obtained analytical expressions allowing  find numerically
the corresponding corrections with any  predetermined accuracy
avoiding lengthy and computer time consuming calculations. With these
analytical expressions in hand, precision of further numerical
calculations is restricted only by the knowledge of the involved
fundamental constants, viz. lepton masses and the fine structure
constant. In the present paper,  we give only the qualitative
peculiarities of the radiative corrections to the anomalies of any type of
leptons from diagrams with  insertions  of one, two and three  lepton loops  in the
polarization operator. An interested reader can easily use the
presented analytical expressions to find numerically the corresponding
corrections with the  desired precision.

As mentioned above,   in the literature
 only corrections   up to the sixth order~\cite{Laporta:1993ju}
were reported in details focusing  mainly on the
 muon anomaly with insertions of electron and muon loops, i.e.
on the sixth order coefficients $A_2^{(6)}(r)$ for $r\, \le\, 1$, in our notation. Analytically, the radiative
corrections for tauons within the considered technique
have not yet been  considered.  Moreover, even for
electrons and muons,  higher order corrections have been investigated
solely as asymptotic expansions. Here  we qualitatively analyze
the fourth, sixth and eighth order coefficients (\ref{A1}) and (\ref{A23}) for all
possible combinations of the existing three   leptons with
one, two and three   internal lepton loops, cf. Eqs.~(\ref{4thCommon}), (\ref{A26LT1}),
(\ref{lLL}), (\ref{Adva}), (\ref{Advaright}),  (\ref{lllLeft}) and
(\ref{Fig3B-D0}).

\begin{figure}[!ht]
\centering
\includegraphics[width=5.4cm,clip]{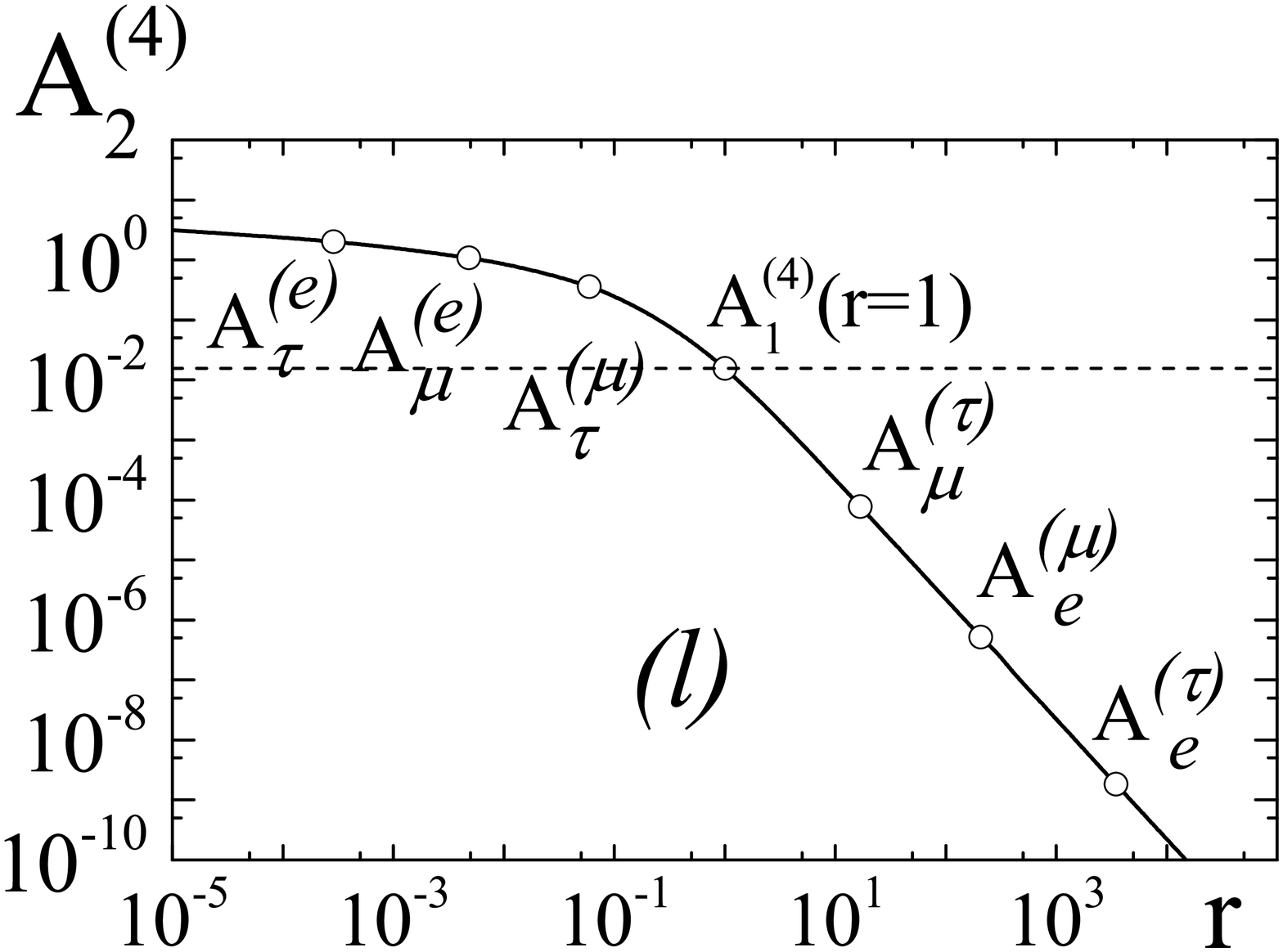}
\includegraphics[width=5.4cm,clip]{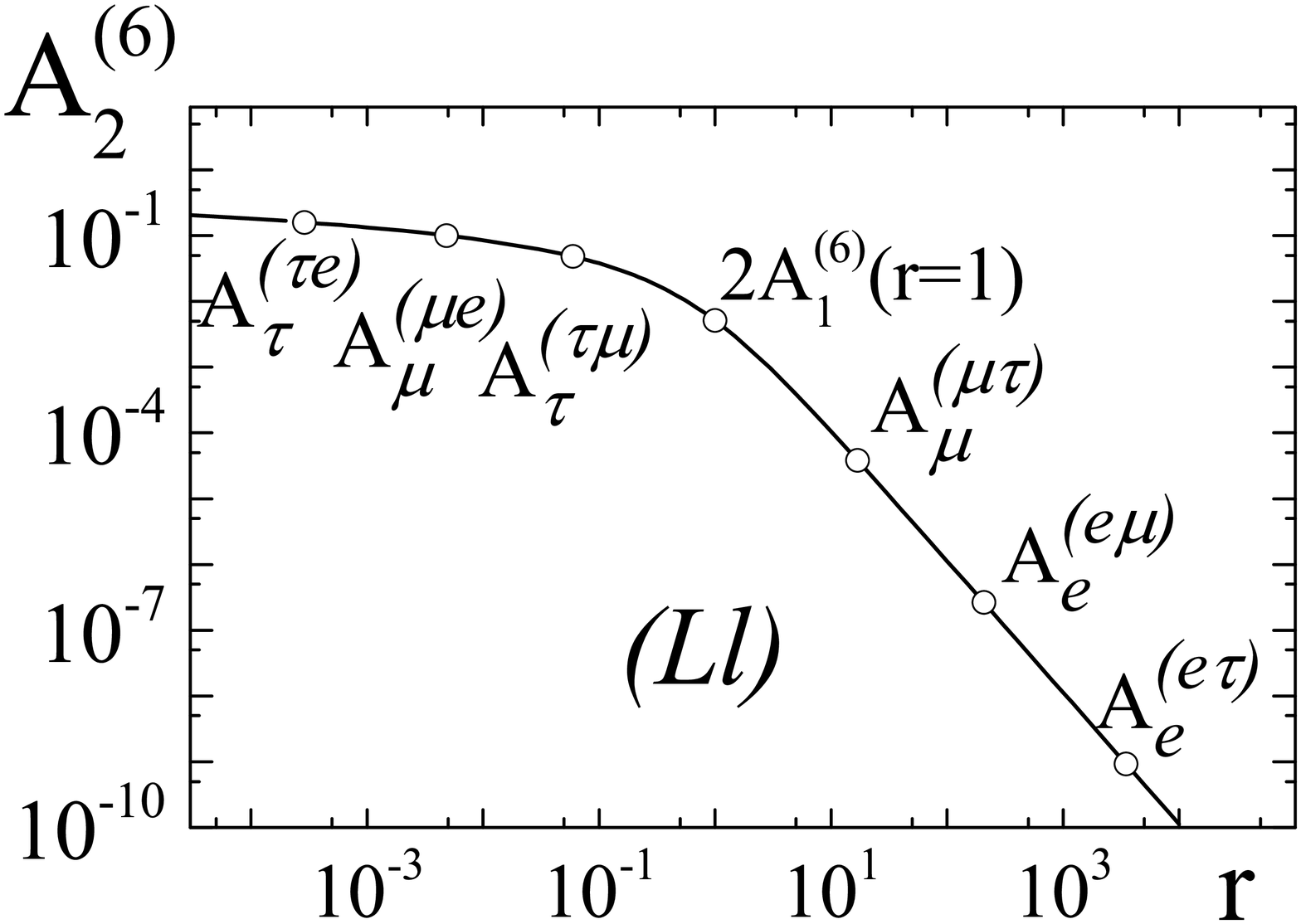}
\includegraphics[width=5.4cm,clip]{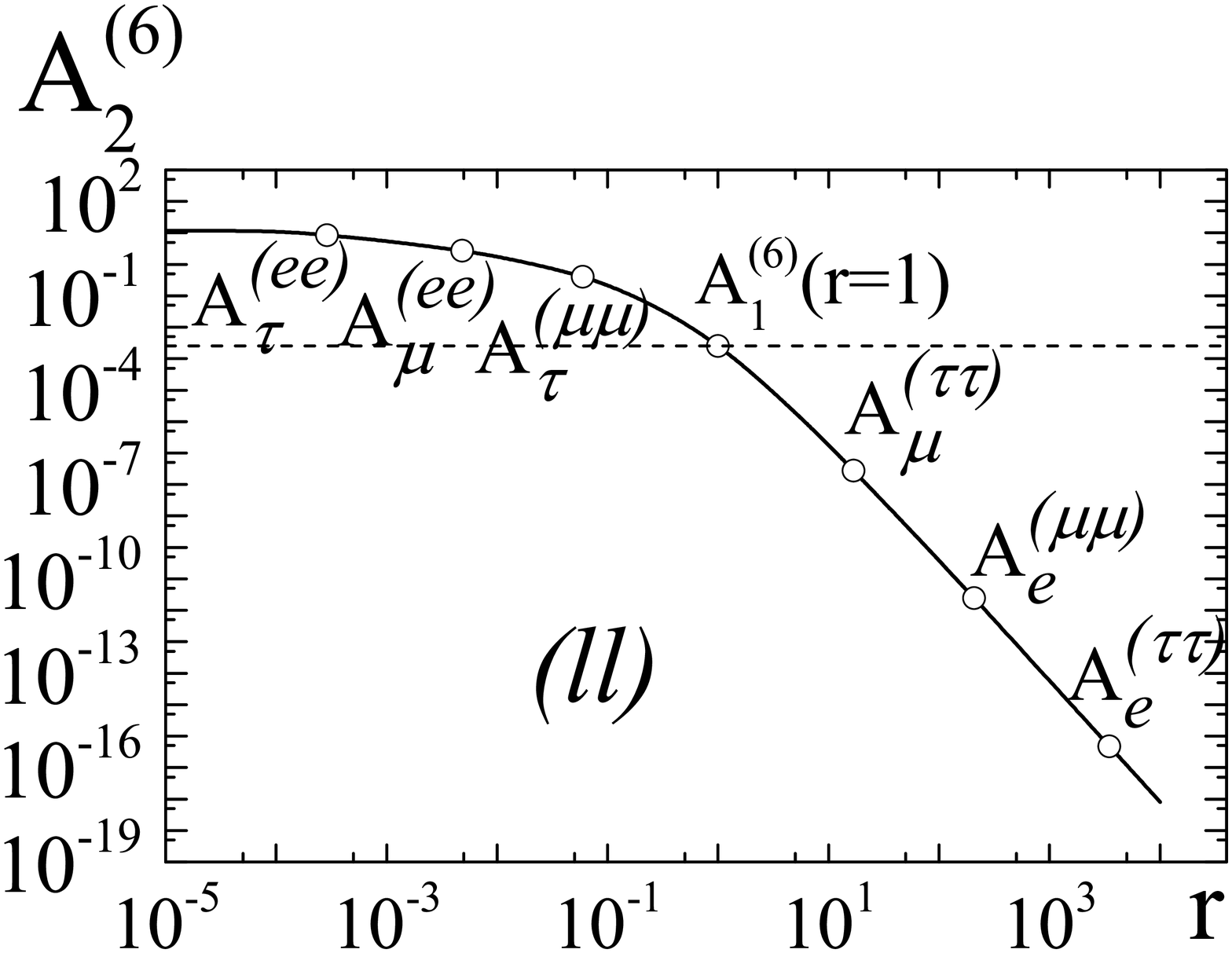}
\caption{The fourth and sixth  order coefficients $A_2^{(4)}(r)$ and  $A_2^{(6)}(r)$, Eq.~(\ref{A23}),
as functions of the mass ratio $r=m_\ell/m_L$,
for an external lepton $L$ with insertions of the polarization operators
with one loop, left panel, and two loops, central and right panels.
The notation corresponds to $A_L^{(\ell)}(r)$,   $A_L^{(L\ell)}$, and $A_L^{(\ell\ell)}$,
 Eqs.~(\ref{4thCommon}), (\ref{lLL}) and (\ref{A26}), respectively, where $L$ and $\ell=e, \mu, \tau$.
The horizontal dashed lines indicate  the values of the coefficients at $r=1$, i.e., the values
$A_1^{(4)}$, Eq.~(\ref{limA1}), and $A_1^{(6)}$, Eq.~(\ref{A16}), which are universal for any kind of the considered leptons.
The central and right panels illustrate the same  dependence of $A_2^{(6)}(r)$  but
   for the combinations $(L \ell)$, Eq.~(\ref{lLL}),
   and $( \ell \ell)$, Eqs.~(\ref{A26}),
respectively. The open circles, as well as the associated with them labels, point to physical values of the ratio $r$ and to the corresponding coefficients $A_2^{(\ell)}(r)$ or $A_2^{(L\ell)}(r)$.}
\label{OneTwoLoops}
\end{figure}
In Fig.~\ref{OneTwoLoops}, we present the results of calculations of the fourth,
   $A_2^{(4)}(r)$,  and sixth order corrections,  $A_2^{(6)}(r)$,
  for diagrams with  one and two lepton loops, as depicted in Fig.~\ref{Fig-2-loop}. The variable $r$ is
defined as $r=m_\ell/m_L$ and varies in the interval $r\in(0,\infty)$.
To ease the visibility of the results, the coefficients $A_2^{(4,6)}(r)$
for   physical values of $r$, i.e., when $L$ and $\ell$ denote the
real existing leptons $e, \mu $ and $\tau$, are marked in
Fig.~\ref{OneTwoLoops} by open circles. Moreover, for a better
illustration  of the combinations of $L$ and $\ell$, each physical
value of $A_2^{(4,6)}(r)$ is additionally labeled by $A_L^{(\ell)} $
(left panel), $A_L^{(L\ell)}$  (central panel) and and
$A_L^{(\ell\ell)}$ (right panel).
Analogously, in Fig.~\ref{Fig-4-QFTHEP} we present the results of calculations of the
eighth order contributions  from diagrams of three types: (i)
insertions of three loops where one loop corresponds to the external
(under consideration) lepton and two loops with lepton pairs different
from the first one, i.e., diagrams of    type $(L\ell\ell)$, left
panel in Figs.~\ref{Fig-3-QFTHEP}~and~\ref{Fig-4-QFTHEP} (ii) one
lepton loop different from the external lepton, two other correspond
to the external one,  i.e., diagrams of the type $(LL\ell)$,  central panel
in Figs.~\ref{Fig-3-QFTHEP}~and~\ref{Fig-4-QFTHEP}, (iii) eventually,
the diagrams with all three loops with leptons of the same type
$(\ell\ell\ell)$, including also the case when $L=\ell$, right panel in
Figs.~\ref{Fig-3-QFTHEP}~and~\ref{Fig-4-QFTHEP}.

\begin{figure}[!ht]
\includegraphics[width=5.4cm,clip]{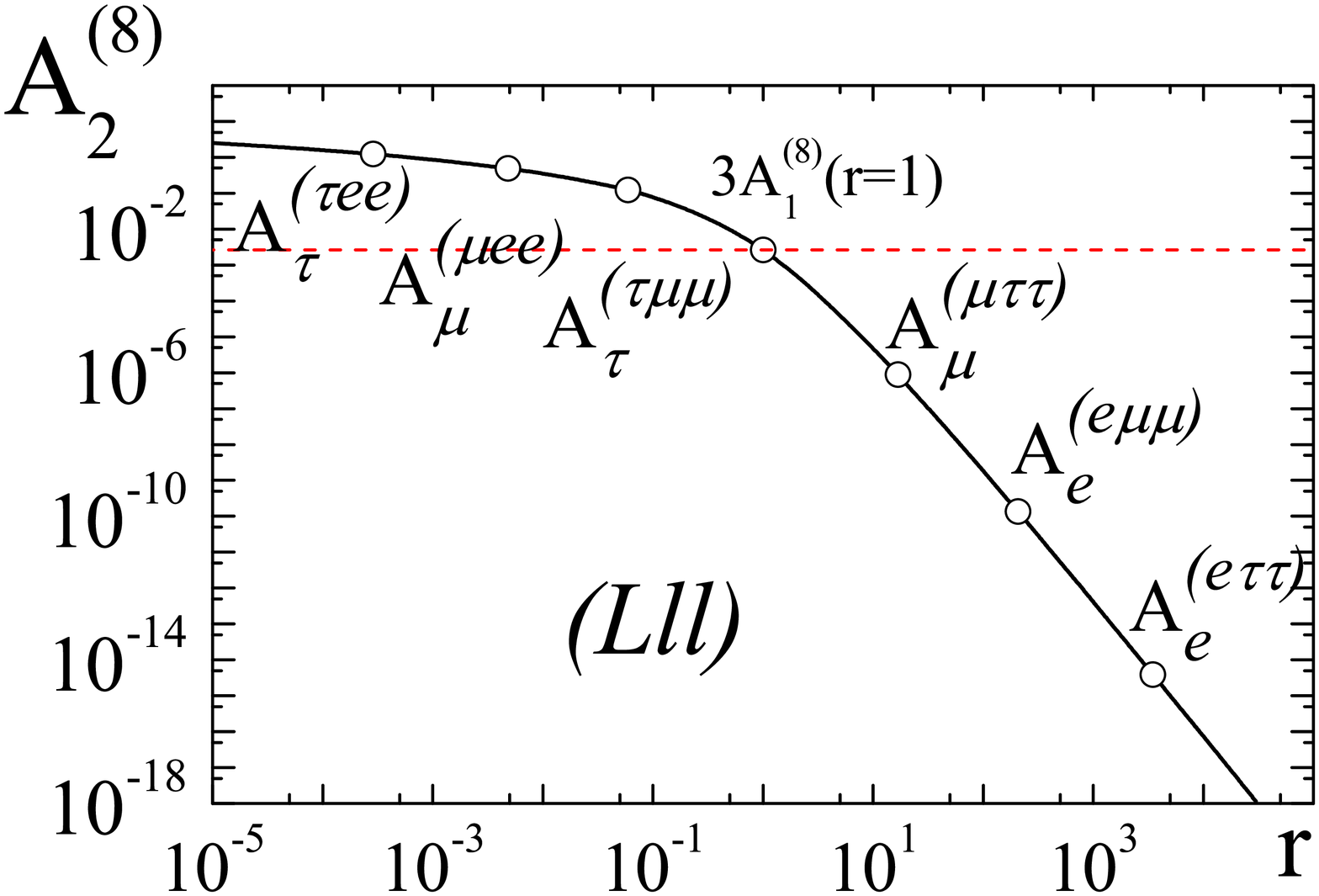}
\includegraphics[width=5.4cm,clip]{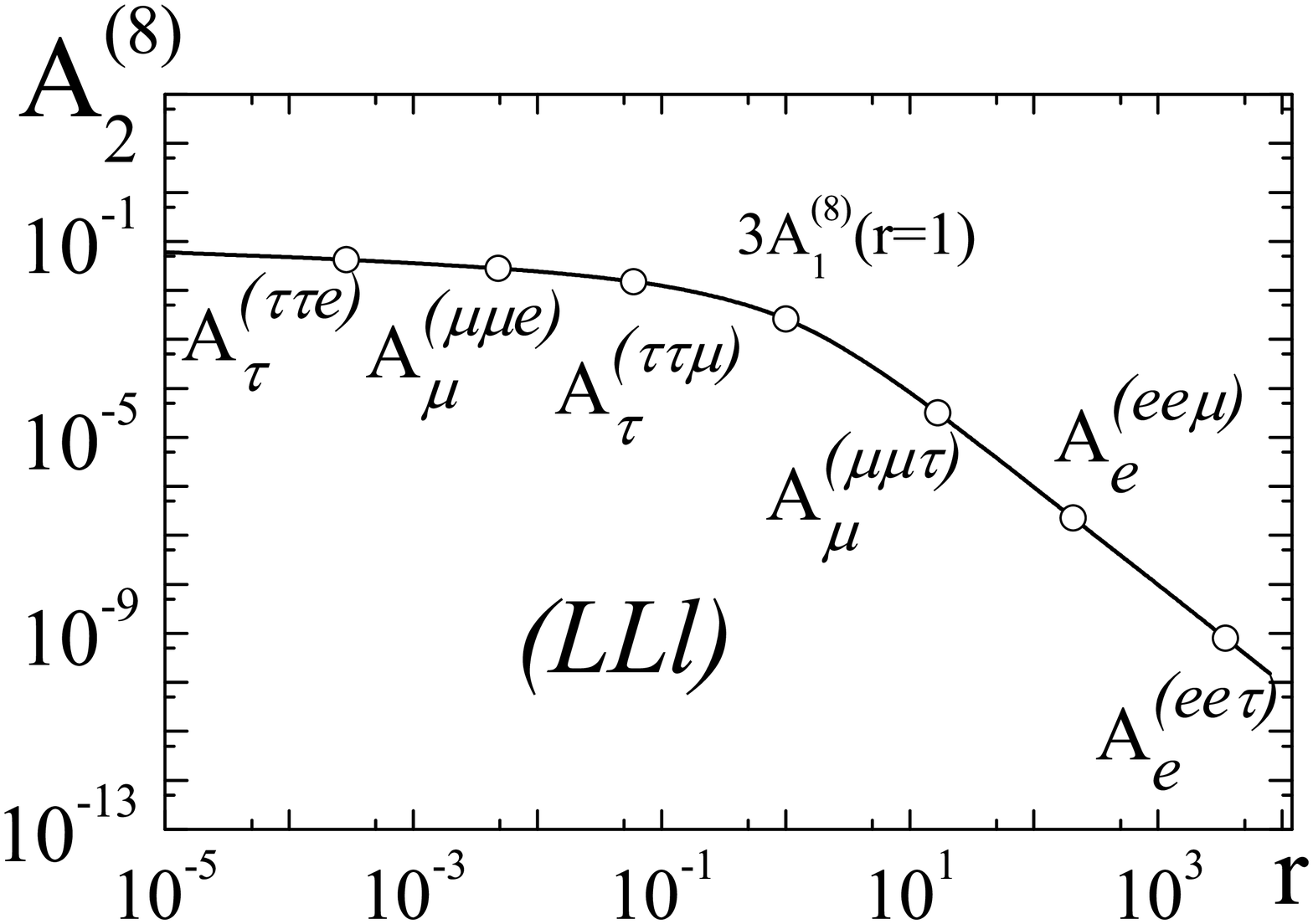}
\includegraphics[width=5.4cm,clip]{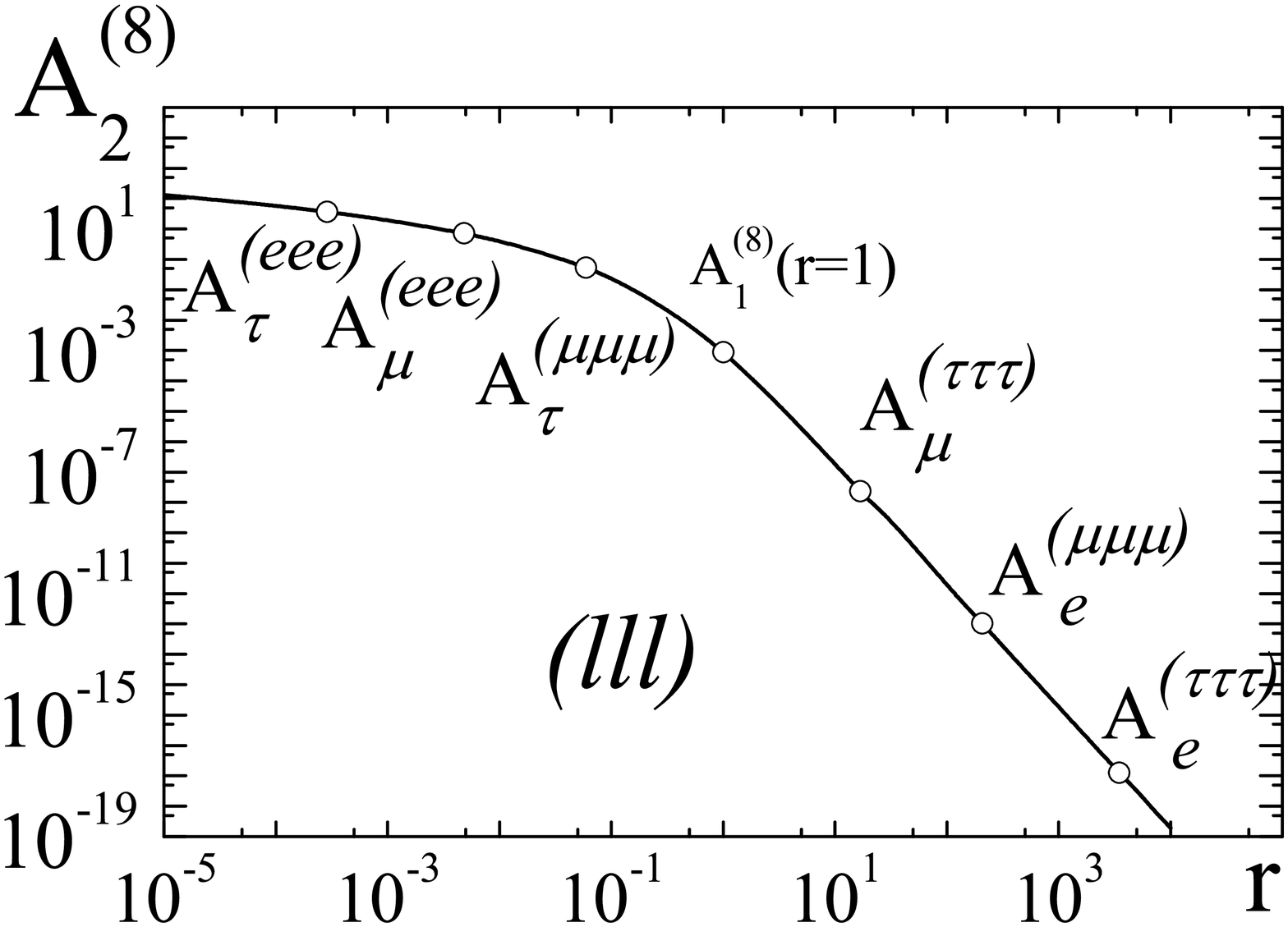}
\caption{The eighth order coefficients $A_2^{(8)}(r)$, Eq.~(\ref{A23}),
as functions of the mass ratio $r=m_\ell/m_L$,
for an external lepton $L=e, \mu, \tau$ with insertions of the polarization operators
with three loops formed by leptons of different types.
Left panel:  the coefficients $A_2^{(8)}(r)$  determined by diagrams with three loops
with one lepton as the external $L$ and two other    different from $L$.
The notation corresponds to $A_L^{(L\ell\ell)}(r)$,  Eq.~(\ref{llL}), where $L$ and $\ell=e, \mu, \tau$.
The horizontal dashed line indicates the value of the coefficients at $r=1$, i.e., the coefficints
$A_1^{(8)}$, Eq.~(\ref{A1}) which are universal for any kind of the considered leptons.
The central and right panels illustrate the same  dependence of $A_L^{(L\ell\ell)}(r)$  but
   for the combinations $(L L \ell)$, Eqs.~(\ref{Adva})-(\ref{sumgt1}),
   and $(\ell \ell \ell)$, Eqs.~(\ref{lllLeft})-(\ref{Fig3B-D0}),
respectively.}
\label{Fig-4-QFTHEP}
\end{figure}

From Figs.~\ref{OneTwoLoops} and \ref{Fig-4-QFTHEP} one infers that the radiative corrections
decrease rather fast,  by about 10-15 orders of magnitude, with
increase of~$r$ from its smallest to highest physical values. It
means that the contribution of the heaviest $\tau$-lepton to the
anomaly of the lighter  ones  is much smaller than the contribution
from properly  the muon  and electron loops. This effect increases with
increasing of   order  of the radiative corrections from $\alpha^2$ to $\alpha^4$ and depends on the combination of
the internal and external leptons.  However, due to the extremely high
precision achieved in the measurements of the lepton
anomalies~\cite{Parker:2018vye,Malaescu,Davoudiasl:2018fbb,E989,E821},
the corrections from $\tau$-leptons cannot be neglected. The horizontal dashed
 lines in Figs.~\ref{OneTwoLoops} and \ref{Fig-4-QFTHEP} correspond
 to  the values of the coefficients $A_2^{(4)-(8)}(r)$ at $r=1$, i.e., to the universal
 coefficients   $A_1^{(4)-(8)}$, Eq.~(\ref{A1}).

Another interesting circumstance to be stressed is that, according to
Eq.~(\ref{fin1}), the radiative corrections are governed not only by
the mass ratio $r$ but, at the same value of $r$, they   are also
quite sensitive to  the combinations of $\Omega_p$ and $R_j$ in
(\ref{fin1}), i.e., to  the concrete form of the  Feynman diagram.
Recall that  $p+j=n$ and that  $\Omega_p$ is determined by the $p-th$
power of the lepton polarization operator, while $R_j$ contains the
imaginary part of the sum $\big [\Pi_l(t) +\Pi_L(t)\big]^j$, cf.
Eqs.~(\ref{Omp}) and (\ref{rho}). So for  the corrections to the
electron anomaly  at, e.g. $r=m_\mu/m_e$, one has $A_e^{ee\mu}(r)\gg
A_e^{e\mu\mu}(r)\gg A_e^{\mu\mu\mu}(r)$ which vary from
$A_e^{ee\mu}(r)\sim 2\cdot 10^{-7}$ to $A_e^{\mu\mu\mu}(r) \sim 2\cdot
10^{-12}$. The same situation occurs for $r=m_e/m_\tau$  as well as
for  muons and tauons. Qualitatively, the relative contributions of
the three loop diagrams to the $e, \mu$ and $\tau$-leptons is
illustrated in Fig.~\ref{Fig-5-QFTHEP}, where the corrections from
different types of insertions decrease from left to right.
%%%%%%%%%%%%%%%%%%%%%%% END  %%%%%%%%%%%%%%%%%%%%
\begin{figure}[!th]
\centering
\includegraphics[width=17cm,clip]{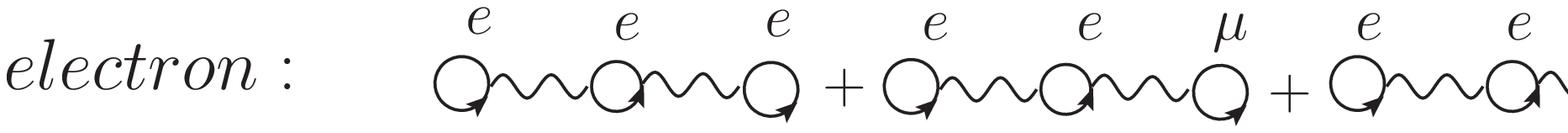}
\includegraphics[width=17cm,clip]{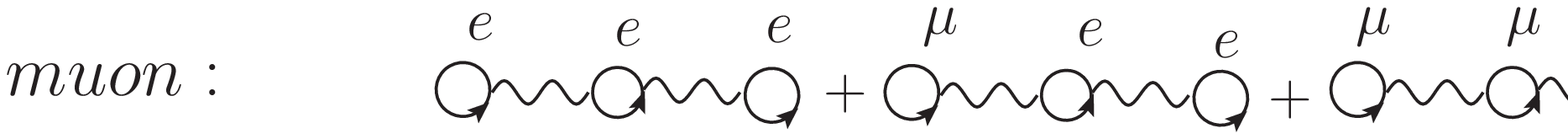}
\includegraphics[width=17cm,clip]{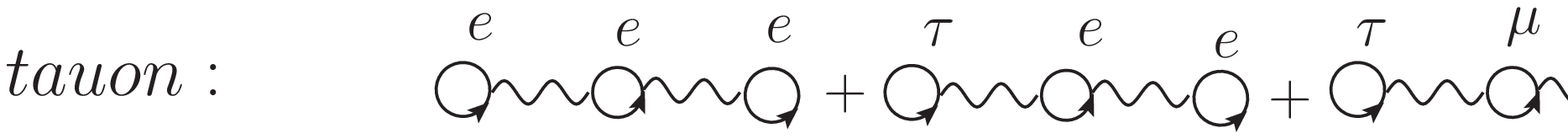}
\caption{Qualitative illustration of the contributions  to the
eighth order coefficients $A_2^{(8)}(r)$, Eq.~(\ref{A23}),
of different types of Feynman diagrams, Fig.~\ref{Fig-3-QFTHEP},  in descending order:
the upper row refers to electrons, the middle row to muons and the lower row to taons.
The notation corresponds to $A_L^{(\ell\ell\ell)}(r)$, $ A_L^{(LL\ell)}(r)$  and $A_L^{(L\ell\ell)}(r)$, where $L=e$ (upper row),
$L=\mu$ (middle row), $L=\tau$ (lower row)
and $\ell=e, \mu, \tau$.}
\label{Fig-5-QFTHEP}
\end{figure}

 Here it is appropriate to reiterate that the  presented results
demonstrate the qualitative analysis of the  contribution of one, two and
three loop  diagrams  to the lepton anomaly. For a scrupulous
quantitative investigation  of the considered radiative corrections
 one can use  the exact analytical expressions reported above.

\subsection{The asymptotic expansions}\label{Subsec:asympt}

The performed  qualitative analysis can be  be essentially  relieved
if, instead of the exact analytical formulae, one employs  their
asymptotic expansions. Such analyzes have been widely used in the
literature \cite{Aguilar:2008qj} by approximate calculations  of the
corresponding integrals (\ref{fin1}) with preliminarily found
asymptotic expansions of  $\Omega_p(s)$ and $R_j(s)$. Explicitly, the
corresponding  expansions  can be found in
Refs.~\cite{Aguilar:2008qj}. In our case, as an additional check  of
the obtained analytical expressions,    we compare our asymptotical
expansions  with the known results reported before  by taking the
limits  $r\ll 1$ and $r\gg 1$ for the coefficients (\ref{A23}). We have
found that   $ A_2^{(8),(L\ell\ell)}(r\ll 1)$, Eq.~(\ref{llL}), $
A_2^{(8),(LL\ell)}(r\ll 1)$, Eq.~(\ref{Adva}), and $
A_2^{(8),(\ell\ell\ell)}(r\ll 1)$, Eq.~(\ref{lllLeft}), are entirely
consistent with the corresponding expressions reported in
Refs.~\cite{Aguilar:2008qj,Kurz:2016bau}. For the sake of brevity, we
do not present them here, mentioning only that such expansions work
surprisingly well in a large interval of $r<1$, from $r\sim 0$  up to
$r\sim 0.3-0.4$.

As for $r\gg 1$, the asymptotic expansion for $A_2^{(8)}(r)$ has not been considered insofar.
 For this reason  below we write out explicitly our  asymptotics,
$r\gg 1$, of the coefficients $ A_2^{(8),(L\ell\ell)}(r\gg 1)$, $
A_2^{(8),(LL\ell)}(r\gg 1)$  and $ A_2^{(8),(\ell\ell\ell)}(r\gg 1)$,
cf. Eqs.~(\ref{llL}), (\ref{Advaright}) and (\ref{Fig3B-D0}):

\noindent
\begin{eqnarray} \label{asymp1}
&&  A_{2}^{(8),(L\ell\ell)}(r\gg 1)
=\left[\frac{5809}{1080000}-\frac{61}{27000}\ln(r)+\frac{2}{225}\ln^2(r)  \right]\frac{1}{r^4}  + \\ &&
 \left[\frac{1862387}{277830000}-\frac{6073}{496125}\ln(r)+
\frac{2}{175}\ln^2(r)\right]\frac{1}{r^6}+ ~~~~~ ~~~~~~~~~~ \nonumber \\ &&
\left[ \frac{12916049}{9001692000}-\frac{1940611}{100018800}\ln(r)+\frac{ 671}{7938}\ln^2(r)\right]\frac{1}{r^8}
+{\cal O}\left(\frac{1}{r^{10}}\right)\, , \nonumber
\end{eqnarray}

\noindent
\begin{eqnarray} &&
   A_{2}^{(8),(LL\ell)}(r\gg 1)   =
 \bigg[\frac{16}{45}\,\zeta (3)  - \frac{203}{486} \bigg]\frac{1}{r^2}+
   \bigg[ \frac {17}{105}\zeta( 3) - \frac {40783}{2315250}\ln ( r)  - \frac{1023526159}{5186160000}
   \nonumber \\ &&-\frac {37}{11025} \ln^2( r)-\frac {2}{315}  \ln^3( r)  \bigg]\frac{1}{r^4}+
   \bigg[ -{\frac {
1243103\, }{93767625}}\ln \left( r \right) -{\frac{4744350631}{
472588830000}}+{\frac {8 }{945}}\,\zeta \left( 3 \right) +\nonumber \\ &&
{\frac {2122\,}{297675}}\ln^2  \left( r \right)-
{\frac {16\,}{2835}}  \ln^3  \left( r \right)  \bigg]  \frac {1}{{r}^{6}}
+\bigg[ \frac {8}{2079} \zeta(3) -\frac{1013327141}{99843767100}\ln(r) -\nonumber \\ &&
 {\frac{8721404003611}{2767669224012000}}+{\frac {166657}{14407470}}\ln^2(r)-
{\frac {16}{6237}}\ln^3(r) \bigg]\frac{1}{r^8}+{\cal O}\left(\frac{1}{r^{10}}\right) \, , \label{asymp2} \\[5mm]
 %\end{eqnarray}
%
%
 %\begin{eqnarray}
    A_{2}^{(8),(\ell\ell\ell)} && (r\gg 1) =
\bigg[\frac{87709}{9729720}-\frac{89}{15015}\zeta(3)\bigg ] \frac{1}{r^4}+
\bigg[\frac{12204667}{1824322500} - \frac{4}{1125}\ln(r)-\frac{40}{9009}\zeta(3)
 \bigg]\frac{1}{r^6} \nonumber \\ &&
+\bigg[\frac{73879547}{17721990000} -\frac{2}{375}\ln(r)-\frac{334}{109395}\zeta(3)\bigg]\frac{1}{r^8}
+{\cal O}\left(\frac{1}{r^{10}}\right) \, .
\label{asymp3}
 \end{eqnarray}
 NB: the asymptotics of the infinite sums $S_2(r)$,
 Eq.~(\ref{sumgt1}), and $S_4(r)$, Eq.~(\ref{s4lg1}), which, as
 mentioned, converge rapidly as $n\gg 1$, have been obtained by
 restricting the summation index $n$ to a finite value $N\sim 5000$
 and then by taking the asymptotics  $r\gg 1$.

Comparing the asymptotic expansion (\ref{asymp1})-(\ref{asymp3}) with
their corresponding exact expressions one concludes that
(\ref{asymp1})-(\ref{asymp3}) provide basically the same (numerical)
results as the exact ones already starting from $r\sim 2 \div 2.5 \,$.

It is worth mentioning that  in computing integrals of   type
(\ref{aL3}), one obtains, depending on the method of integration used
in dependence of the used,  analytical results in terms of various special
functions, viz. polygammas $\psi^{(n)}(r)$, polylogarithms ${\rm
Li}_n(r)$, harmonic functions $H(n,r)$, Hurwits-Lerch transcendent
$\Phi(r,n,a)$ etc. To reconcile different analytical  results to each
other,  the number of   special functions involved in integration
must be maximally reduced by using the known relations among  these
special functions (see Appendix~\ref{app}). In such a way, the
asymptotical expressions for $r\ll 1$ have been
identically reduced  to the previously reported expansions. This
further  persuades us of the correctness of our analytical analysis of the
coefficients (\ref{A1})-(\ref{A23}) and, consequently,   that  for each
coefficient in Eq~(\ref{A23})  there is a corresponding analytical
function valid in the whole interval $r\in (0,\infty)$.

 \section{Summary} \label{summary}

In summary,   we have presented an investigation of the
contributions to the anomalous magnetic moment of leptons
 $L$  ($L=e\, ,\,\mu $ or  $\tau$) generated by   QED diagrams with
insertions of one, two and three loops in the photon vacuum  polarization
operator. We considered all possible combinations of external   and
internal leptons in the bubble-like diagrams. The radiative
corrections for each  diagram  are obtained in close analytical
forms. We argued that  each coefficient  $A_2(r)$ determining the
corresponding $4$th, $6 $th and $8 $th order of the radiative
corrections can be represented explicitly by analytic functions which
depend on the mass ratio $r=m_l/m_L$,   being different for different
combinations  of the external $(L)$ and internal ($\ell$)  leptons  in
the Feynman diagrams of the corresponding order.  The generic variable
$r$ of these functions is  defined in the whole region $0 < r < \infty$.

Our consideration is based on a combined use of the dispersion
relations for the vacuum polarization operators and the Mellin-Barnes
integral transform for the Feynman parametric integrals. This  technique is widely
used in the literature in multi-loop  calculations
in relativistic quantum field theories,  c.f.~Refs.~\cite{Friot:2005cu,Rafael-HVP,Kotikov:2018wxe}.  The
ultimate integrations have been performed by the Cauchy residue theorem in
the left ($r<1$) and right ($r>1$) semiplanes of the complex Mellin
variable $s$. We demonstrate that the results in these two regions
complement each other and determine the two branches of
 a common, for each considered Feynman diagram, analytical function.
  We investigated numerically the
behaviour of these functions in the whole interval  $0 < r < \infty$
and   classified for each lepton  the contribution of various diagrams
in descending order of their  significance.  We also showed that the diagrams
with $r>1$, i.e. diagrams with loops formed by heavier leptons, play a
minuscule role in comparison with the role of corrections in the
interval $r<1$. However, in high precision calculations such
contributions can not be neglected.

Whenever pertinent, we compared our analytical expressions and  the
corresponding asymptotical  expansions with  well-known results available
in the literature and found that they are fully compatible with the
early known calculations.

The present paper can be considered as further developments of the
efforts aimed at a better understanding of  the role of bubble-like diagrams
in the lepton anomaly and as an extension of previously reported analysis to the
all three muons in the whole interval of the mass ratio $r$, $(0\, < \, m_l/m_L\, < \,\infty)$.  The performed analysis persuades us that the Mellin-Barnes approach probably  can be successfully applied in calculating
analogous diagrams with insertions, this time,  of  hadron loops in
the polarization operator.

 \section* {\bf Acknowledgments} \vskip 0.2cm

We gratefully acknowledge helpful discussions with  A.~L.~Kataev and
O.~V. Teryaev  and  their support of the present  activity. We also
thank A.~V.~Sidorov for discussions and  cooperation in the earlier
stages of this work.

\appendix

 \section{Some useful relations}\label{app}
Direct employment  of the Cauchy residue theorem to the integrals,  Eqs.~(\ref{fin23}), (\ref{A26}),
(\ref{fin3-differ}), (\ref{aLtwo}), (\ref{aL3}), and (\ref{Fig3}) results in expressions containing a variety
of special functions, viz. polygammas $\psi^{(m)}(n)$, polylogarithms ${\rm
Li}_n(r)$, generalized harmonic functions $H_n^{(m)}$, Hurwits-Lerch transcendent
$\Phi(r,n,a)$ etc. Using the herebelow relations,
  the  number of necessary special functions can be essentially reduced. This allows one to reconcile explicitly
 our results  to the ones known in the literature and reported in different
  forms with different special functions, cf. Refs.~\cite{Friot:2005cu,Laporta:1993ju,Sidorov:2019}.
 Also, by using the appropriate  properties of the remaining functions for $r<1$ and $r>1$, one can
 express the corresponding coefficients $A_2(r<1)$ through the coefficients $A_2(r<1)$, which allows one to assert that  there exist, for each Feynman diagram, a common analytical function valid in the whole interval $(0\,<\,r\,\infty)$.

  \ba &&
\begin{array}{lr}
\Phi(r,2,1/2) = \frac{2}{\sqrt{r}} \left[ \li2 (\sqrt{r}) - \li2 (-\sqrt{r}) \right] ;     \\[3mm]
\Phi(r,2,3/2) =-\frac4r +\frac{2}{r\sqrt{r}} \left[ \li2 (\sqrt{r}) - \li2 (-\sqrt{r}) \right];     \\[3mm]
\Phi(r,2,5/2) =-\frac{4}{9r} - \frac{4}{r^2} +\frac{2}{r^2\sqrt{r}} \left[ \li2 (\sqrt{r}) - \li2 (-\sqrt{r}) \right];       \\[3mm]
 \li2 (1-r)+ \li2 \left(1-\frac1r\right) =-\frac12 \ln^2(r)   ;  \qquad  \qquad         {\rm (r>0)}; \\[3mm]
 \li2 (r) +  \li2 (1/r)  = -\frac{\pi^2}{6} - \frac12  \ln^2(-r) ; \qquad \qquad   {\rm (r>1)}; \\[3mm]
\Li3 (r) - \Li3 (1/r) = -\frac{\pi^2}{6}\ln(-r)  - \frac16 \ln^3(-r)  ;  \qquad   {\rm (r>1)} \\[3mm]
\Li3 (r) - \Li3 (1/r) = \frac{\pi^2}{6}\ln(-1/r)  + \frac16 \ln^3(-1/r)  ; \qquad    {\rm (r<1)} \\[3mm]
\poly4 (r) + \poly4 (1/r) =-\frac{7\pi^4}{360} -\frac{\pi^2}{12}(\ln(-r))^2   -
\frac{1}{24} \ln^4(-r)  ;
\qquad  \qquad  {\rm (r>1)}
\\[3mm]
 {\rm Li_2} \left( \frac{1-r}{1+r} \right)-
 {\rm Li_2}\left( - \frac{1-r}{1+r} \right)= {\rm Li_2}(-r) - {\rm Li_2}(r)
 +\bigg(\ln(1+r)\ln(r) - \ln(1-r)\bigg)\ln(r) + \frac{\pi^2}{4};       \\[5mm]
{\rm Li_n}(r) + {\rm Li_n}(-r) = \frac{1}{2^{n-1}} {\rm Li_n}(r^2).    \\[3mm]
\end{array}
\\ &&
{ \rm arctanh(r)}=\frac12\bigg [ \ln(1+r)-\ln(1-r)\bigg ]; \\ &&
  {\rm H_n^{(1)}} = \psi^{(1)}(n+1)+\gamma;  \quad {\rm H}_n^{(2)}=\frac{\pi^2}{6}-\psi^{(1)}(n+1),
  \ea
  where ${\rm Li_n}(r)$, $\Phi(r,s,a) $, ${\rm H_n^{(m)}} $  and $\psi^{(m)}(n)$ are the polylogaritm, Lerch transcendent,
  generalized harmonic number and Euler polygamma functions, respectively.

\end{document}